# Light emission from strongly driven many-body systems

Andrea Pizzi[1,2,†], Alexey Gorlach[3,†], Nicholas Rivera[1,4], Andreas Nunnenkamp[5], and Ido Kaminer[3]

[1]*Department of Physics, Harvard University, Cambridge 02138, Massachusetts, USA*
[2]*Cavendish Laboratory, University of Cambridge, Cambridge CB3 0HE, United Kingdom*
[3]*Solid State Institute, Technion–Israel Institute of Technology, Haifa 32000, Israel*
[4]*Massachusetts Institute of Technology, 77 Massachusetts Avenue, Cambridge 02139, Massachusetts, USA*
[5]*Faculty of Physics, University of Vienna, Boltzmanngasse 5,1090 Vienna, Austria*
*Corresponding authors: I.K. ([kaminer@technion.ac.il](kaminer@technion.ac.il)) and N.R. ([nrivera@mit.edu](nrivera@mit.edu))*
*† equal contribution*

**Strongly driven systems of emitters offer an attractive source of light over broad spectral ranges up to the X-ray region. A key limitation of these systems is that the light they emit is for the most part classical. We challenge this paradigm by building a quantum-optical theory of strongly driven many-body systems, showing that the presence of correlations among the emitters creates emission of nonclassical many-photon states of light. We consider the example of high-harmonic generation (HHG), by which a strongly driven system emits photons at integer multiples of the drive frequency. In the conventional case of uncorrelated emitters, the harmonics are in an almost perfectly multi-mode coherent state lacking any correlation between harmonics. By contrast, a correlation of the emitters prior to the strong drive is converted onto nonclassical features of the output light, including doubly-peaked photon statistics, ring-shaped Wigner functions, and quantum correlations between harmonics. We propose schemes for implementing these concepts – creating the correlations between emitters via an interaction between them or their joint interaction with the background electromagnetic field (as in superradiance). By tuning the time at which these processes are interrupted by the strong drive, one can control the amount of correlations between the emitters, and correspondingly the deviation of the emitted light from a classical state. Our work paves the way towards the engineering of novel many-photon states of light over a broadband spectrum of frequencies, and suggests HHG as a diagnostic tool for characterizing correlations in many-body systems with attosecond temporal resolution.**

# Introduction

The creation and control of many-photon quantum states of light are important problems with applications across the natural sciences. Realizations of squeezed quantum light states open new avenues in spectroscopy and metrology, providing novel information on samples [1] and enabling highly sensitive measurements beyond classical noise limits (e.g. in the detection of gravitational waves [2,3]). At the same time, encoding quantum information on the quantum state of light facilitates applications in quantum computing, simulation, and communication [4]. Several pioneering investigations have demonstrated a range of many-photon quantum states of light, such as squeezed light [2,3,5–7], bright squeezed vacuum [8–11], displaced Fock states [12], Schrodinger kitten [13,14] and cat states [15,16], subtracted squeezed states [17], and others [18]. Many of the established techniques for generating quantum light at optical frequencies rely on materials with a nonlinear optical response. Such nonlinear materials can be typically described using a "perturbative" nonlinear response, where the induced polarization is for example quadratic or cubic in the applied electric field.

At the other extreme of nonlinear optics are "non-perturbative" or "strong-field" effects like high-harmonic generation (HHG), in which a very intense optical pulse creates radiation at very high frequencies, even beyond hundredfold the frequency of the drive [19,20]. As such, HHG is an attractive source of ultra-short pulses of high-frequency light. The potential of HHG for the generation of nonclassical high-frequency light has, however, remained largely unexpressed. In fact, many past semi-classical approaches [21–23] and more recent fully quantized ones [14,24] have established that the output harmonics in HHG are in an almost precisely coherent (thus, classical) state (apart from the notable exception of post-selected cat states in the driving frequency [14]). Nonetheless, these works all focus on the scenario of uncorrelated emitters, leaving open important questions about many-body aspects underlying HHG. In particular, to what extent do correlations between the emitters affect the state of light created in the HHG process?

In this work, we develop a quantum-optical theory of light emission by strongly driven many-body systems. We use this theory to show that many-body correlations in the emitters can render the output radiation strongly nonclassical (Figure 1). To demonstrate this concept, we show that HHG from a correlated many-body state of emitters features exotic photonic states, for instance, super-Poissonian and doubly peaked photon number statistics, ring-shaped Wigner functions, and strong correlations between harmonics. These features strongly contrast with conventional HHG from uncorrelated emitters, in which the output harmonics are described by almost-perfectly Poissonian photon statistics, Gaussian Wigner function, and uncorrelated harmonics.

Indeed, the quantum state of the emitted light can be shaped by creating different correlations among the emitters. We show this general idea by investigating two concrete scenarios, one in which correlations among the emitters are induced through collective superradiant emission, and one in which they arise from dipole-dipole type interactions. Our study makes the first step towards the creation of bright high-frequency light with engineered quantum properties. Applying this concept in reverse, characterizing the quantum photonic state of the emitted light will enable to infer the many-body correlations of the material with high-temporal-resolution.

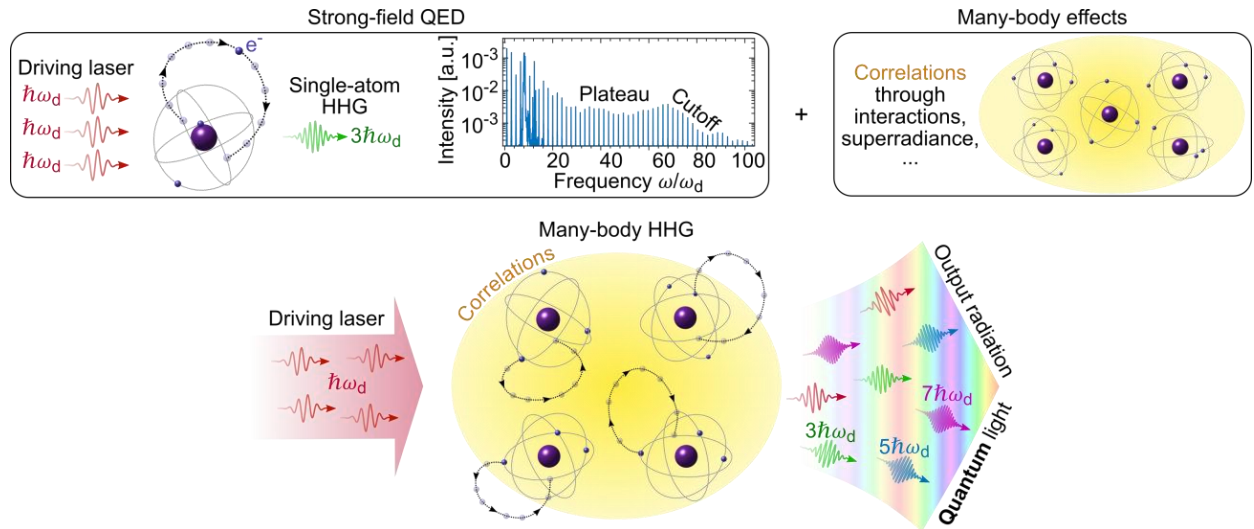

**Figure 1. Quantum theory of light emission by strongly driven many-body atomic systems.** **(a)** HHG can be understood as a single-particle, strong-field three-step process: (i) an intense drive laser tears off an electron from the atom, (ii) the electron is accelerated by the electric field, (iii) the electron recombines with the atom, converting its energy into an energetic photon at higher harmonics. The spectrum features are peaks at the odd harmonics, a characteristic plateau and a cutoff. **(b)** The many-body correlations among the atoms can arise from spontaneous collective emission (superradiance) or interatomic interactions. **(c)** Our theory marries the description of many-body effects with that of strong-field physics, giving access to unique quantum properties of the emitted light.

## Results

We develop a quantum-optical theory that describes the interaction among an intense driving field, a quantum-correlated (many-body) atomic system, and the quantized radiation emitted from it. For a given initial atomic condition, our theory produces a full portrait of the emitted quantum light, including the Wigner function, photon number statistics, and correlation between different harmonics. First, we consider the standard scenario in which the atoms are initially in their ground state and show that the emission resulting from the strong drive is in this case essentially coherent: the Wigner function and photon statistics of the harmonics are almost Gaussian and Poissonian, respectively, and the different harmonics are uncorrelated. Second, we investigate the effect of strong atomic correlations on the output radiation, showing that, when the emitters are initially in a correlated (thus, non-separable) many-body state, the output radiation becomes nonclassical, featuring non-Gaussian Wigner functions, non-Poissonian photon statistics, and correlated harmonic pairs. Third, we propose two experimentally relevant schemes, exploiting inter-atomic interactions and superradiance, respectively, to induce correlations in the atomic initial condition, resulting in controllably nonclassical output radiation. An expanded and detailed version of the following theory is presented in the Supplementary Information (SI). In the following, Hartree atomic units ($m_e = \hbar = e = a_0 = 1$, and $c = 137$) are used unless otherwise specified.

### 1 Quantum theory of strongly driven many-body systems

We consider $N$ microscopic emitters (henceforth referred to as 'atoms' for convenience) interacting with a strong driving field. The Hamiltonian describing the system reads

$$\widehat{H} = \sum_i^N \widehat{H}_{\text{atom}}^i + \sum_i^N \widehat{\boldsymbol{d}}_i \cdot \widehat{\boldsymbol{E}}(\boldsymbol{r}_i) + \widehat{H}_{\text{F}}, \tag{1}$$

where $\boldsymbol{r}_i$ and $\widehat{\boldsymbol{d}}_i$ are the position and dipole moment of the $i$-th atom, respectively, $\widehat{H}_{\text{F}} = \sum_{k\sigma} \hbar\omega_k \hat{a}_{k\sigma}^\dagger \hat{a}_{k\sigma}$ is the free-field Hamiltonian, with $\hat{a}_{k\sigma}^\dagger$ and $\hat{a}_{k\sigma}$ the creation and annihilation operators of a photon with wavevector $\boldsymbol{k}$ and polarization $\sigma$, respectively, and $\widehat{H}_{\text{atom}}^i$ is the single particle Hamiltonian describing the outermost electron of the $i^{\text{th}}$ atom. The eigenvalue problem $\widehat{H}_{\text{atom}}^i |m\rangle = w_m |m\rangle$ is solved via standard space discretization procedures, resulting in a discrete single-particle spectrum composed of hundreds of levels, most of which in the continuum, $w_m > 0$ (see Supplementary Section S1).

The atomic system is driven with a laser whose field we take as a multimode coherent state $|\psi_{\text{laser}}(t)\rangle = \prod_{\boldsymbol{k}}|\alpha_{\boldsymbol{k}}e^{i\omega_{\boldsymbol{k}}t}\rangle$, where $\alpha_{\boldsymbol{k}}$ is the parameter of a coherent state with frequency $\omega_{\boldsymbol{k}}$. Using a unitary transformation (generated by a displacement operator) [24–28], the electric field $\widehat{\boldsymbol{E}}(\boldsymbol{r})$ can be separated into a classical part $\boldsymbol{E}_{\text{c}}(\boldsymbol{r},t) = \langle\psi_{\text{laser}}(t)|\widehat{\boldsymbol{E}}(\boldsymbol{r})|\psi_{\text{laser}}(t)\rangle$ and a quantum part $\widehat{\boldsymbol{E}}_{\text{q}}(\boldsymbol{r}) = -\sum_{\boldsymbol{k}\sigma} g_{\boldsymbol{k}\sigma}\boldsymbol{\varepsilon}_{\boldsymbol{k}\sigma}\hat{a}_{\boldsymbol{k}\sigma} + h.c.$, where $g_{\boldsymbol{k}\sigma} = i\sqrt{\frac{2\pi\omega_{\boldsymbol{k}}}{V}}$, $\boldsymbol{\varepsilon}_{\boldsymbol{k}\sigma}$ is the polarization vector of the mode $\boldsymbol{k}\sigma$, and $V$ is the volume. The term $\widehat{\boldsymbol{E}}_{\text{q}}$ represents quantum fluctuations of the electric field. We are interested in describing emission in vacuum, for which $V \to \infty$ and the modes are continuous. Under the action of the displacement operator, the initial photonic state changes from $|\psi_{\text{laser}}\rangle$ to the vacuum with zero photons $|0\rangle$, and the photonic state only describes the radiation on top of the drive laser. Assuming the electric field $\widehat{\boldsymbol{E}}(\boldsymbol{r})$ to be polarized the $x$ direction, the light-matter coupling term in Eq. (1) can be rewritten as $\sum_i^N \hat{x}_i \hat{E}(\boldsymbol{r}_i)$. In its single-particle version with $N = 1$ atoms, the above model is known to capture the main features of HHG, including the characteristic plateau and cutoff of the emission spectrum [21,22], see Figure 1b. Here, we wish to investigate the effects that many-body atomic correlations for $N \gg 1$ have on the emitted output light. Due to the exponential size (in $N$) of the Hilbert space, such a many-body problem is in general a formidable one. We make it tractable by considering the simplest yet far from trivial scenario, one in which the spatial arrangement of the atoms is negligible. In this case, the state of the system can remain symmetric, meaning invariant under any permutation of the atoms, thus making the atoms effectively indistinguishable. While the assumption of indistinguishable atoms is physically justified in some cases, e.g., when the atoms are in a small volume with respect to the involved interaction ranges and wavelengths, it can qualitatively describe multiple many-body phenomena even when not fully justified, similarly to what happens for instance for superradiance [29,30], spin squeezing [31], and countless equilibrium phase transitions [32,33]. Indeed, this assumption opens the way to much analytical progress, which brings the ultimate numerical simulation of the system into reach and greatly facilitates physical intuition on the core involved physics.

The first key step is to note that, thanks to the atoms' permutational symmetry above, and according to a procedure analogous to second quantization, the state of the atomic system can be expressed in terms of atomic Fock states $|\boldsymbol{n}\rangle = |n_1, n_2, ..., n_m, ...\rangle$ with $n_m$ the number of atoms in the $m$-th single particle level of $H_{\text{atom}}^i$ (Supplementary Section S2). In terms of standard bosonic creation and annihilation operators $\hat{b}_m^\dagger$ and $\hat{b}_m$, these states read $|\mathbf{n}\rangle = \left(\prod_m (\hat{b}_m^\dagger)^{n_m}/\sqrt{n_m!}\right)|0\rangle$, whereas the Hamiltonian (1) yields

$$H = \hat{\mathbf{b}}^\dagger \mathbf{W}\hat{\mathbf{b}} - \left(E_c(t) + \hat{E}_q\right)\hat{\mathbf{b}}^\dagger \mathbf{D}\hat{\mathbf{b}} + \sum_{\mathbf{k}\sigma} \omega_k \hat{a}_{\mathbf{k}\sigma}^\dagger \hat{a}_{\mathbf{k}\sigma}, \qquad (2)$$

where $\mathbf{W}$ is a diagonal matrix with single-atom energies as entries, namely $W_{mn} = \delta_{mn} w_m$, $\mathbf{D}$ is a dipole matrix with entries $D_{mn} = \langle m|\hat{x}|n\rangle$, and where we called $\hat{\mathbf{b}} = (\hat{b}_1; \hat{b}_2; \dots)$ the column vector of bosonic annihilation operators and $\hat{\mathbf{b}}^\dagger = (\hat{b}_1^\dagger, \hat{b}_2^\dagger, \dots)$ the row vector of bosonic creation operators. In particular, the first two modes, $\hat{b}_1$ and $\hat{b}_2$, refer to the single particle ground and first excited states $|g\rangle$ and $|e\rangle$, respectively. The Heisenberg equations associated with Eq.(2) read

$$\begin{cases} \dfrac{d\hat{\mathbf{b}}}{dt} = -i\left(\mathbf{W} - \left(E_c(t) + \hat{E}_q\right)\mathbf{D}\right)\hat{\mathbf{b}}, \\ \dfrac{d\hat{a}_{\mathbf{k}\sigma}}{dt} = -i\omega_k \hat{a}_{\mathbf{k}\sigma} + i g_{\mathbf{k}\sigma}^* \boldsymbol{\varepsilon}_{\mathbf{k}\sigma} \hat{\mathbf{b}}^\dagger \mathbf{D}\hat{\mathbf{b}}. \end{cases} \qquad (3)$$

To solve Eq. (3) we neglect the quantized part of the field $\hat{E}_q$ in favor of the strong classical drive $E_c$, as customary for HHG [22,34–36]. Indeed, this makes the atomic equation linear, which allows us to write

$$\hat{\mathbf{b}}(t) = \mathbf{F}(t)\hat{\mathbf{b}}(0), \quad \mathbf{F}(t) = T\exp\left(-i\int_0^t (\mathbf{W} - E_c(\tau)\mathbf{D})d\tau\right), \qquad (4)$$

a time evolution matrix, and $T$ denoting time-ordering. As for the photons, integrating the second equation in Eq. (3), and plugging Eq. (4) in, we get

$$\hat{a}_{\mathbf{k}\sigma}(t) = \hat{a}_{\mathbf{k}\sigma}(0) + i g_{\mathbf{k}\sigma}^* \boldsymbol{\varepsilon}_{\mathbf{k}\sigma} \hat{\mathbf{b}}^\dagger(0)\tilde{\mathbf{D}}(\omega_k)\hat{\mathbf{b}}(0). \qquad (5)$$

where $\tilde{\mathbf{D}}(\omega) = \int_0^t d\tau e^{i\omega(\tau-t)} \mathbf{F}^\dagger(\tau)\mathbf{D}\mathbf{F}(\tau)$. Considering only initial conditions with atoms in the two lowest single particle states $|g\rangle$ and $|e\rangle$, we can effectively replace $\hat{\mathbf{b}}^\dagger(0) \to \left(\hat{b}_1^\dagger(0), \hat{b}_2^\dagger(0)\right)$, $\hat{\mathbf{b}}(0) \to \begin{pmatrix} \hat{b}_1(0) \\ \hat{b}_2(0) \end{pmatrix}$, and $\tilde{\mathbf{D}} \to \tilde{\mathbf{d}} = \begin{pmatrix} \tilde{d}_{11} & \tilde{d}_{12} \\ \tilde{d}_{21} & \tilde{d}_{22} \end{pmatrix}$, which will henceforth be implicit in our notation. That is, for this class of initial conditions, the information contained in the $M \times M$-dimensional matrix $\mathbf{F}$, that accounts for all the atomic levels including the continuum, gets compressed into the $2 \times 2$ dimensional matrix $\tilde{\mathbf{d}}$. Next, we introduce the mode $\hat{a}_n$ of the $n^{\text{th}}$ harmonic as a normalized sum $\hat{a}_n = \frac{1}{\sqrt{N}} \sum_{\mathbf{k} \in n} \hat{a}_{\mathbf{k}\sigma}(t_f)$, with $t_f$ the time at which the pulse is over, and running over wavevectors $\mathbf{k}$ within a given solid angle $d\Omega$ and with frequency $\omega_k$ within the range $\left(n\omega_d - \frac{d\omega}{2}, n\omega_d + \frac{d\omega}{2}\right)$, that should be though of as those characterizing a detector used to collect the output light. Enforcing $[\hat{a}_n, \hat{a}_n^\dagger] = 1$ we find $\mathcal{N} = \frac{n^2\omega_d^2 V d\Omega}{(2\pi)^3 c^3} d\omega$, and get

$$\hat{a}_n = \hat{a}_n(0) + \hat{\boldsymbol{b}}^\dagger(0) \boldsymbol{d}_n \hat{\boldsymbol{b}}(0), \tag{6}$$

with $\boldsymbol{d}_n = \sqrt{\frac{d\Omega}{4\pi} \frac{n^3 \omega_d^3}{\pi c^3} d\omega}\, \tilde{\boldsymbol{d}}(n\omega_d)$. We can decompose the matrix $\boldsymbol{d}_n$ in terms of Pauli matrices as $\boldsymbol{d}_n = \alpha + (\boldsymbol{u}_n + i\boldsymbol{v}_n) \cdot \boldsymbol{\sigma}$, where $\alpha = \frac{N}{2}\text{Trace}(\boldsymbol{d}_n)$ is a complex number, $\boldsymbol{u}_n$ and $\boldsymbol{v}_n$ two 3-component real vectors, and $\boldsymbol{\sigma} = (\sigma_x, \sigma_y, \sigma_z)$ the vector of Pauli matrices. We note that $N = \hat{\boldsymbol{b}}^\dagger(0)\hat{\boldsymbol{b}}(0)$ and that $\hat{\boldsymbol{S}} = (\hat{S}_x, \hat{S}_y, \hat{S}_z) = \sum_j^N \hat{\boldsymbol{\sigma}}_j = \hat{\boldsymbol{b}}^\dagger(0)\boldsymbol{\sigma}\hat{\boldsymbol{b}}(0)$, with $\hat{\boldsymbol{\sigma}}_j = (|e\rangle_j\langle g|_j + |g\rangle_j\langle e|_j, i(|g\rangle_j\langle e|_j - |e\rangle_j\langle g|_j), |e\rangle_j\langle e|_j - |g\rangle_j\langle g|_j)$ the Pauli operators describing the transition between each atom's ground and first excited states $|g\rangle$ and $|e\rangle$, respectively. We can thus rewrite Eq. (6) in terms of standard collective spin operators for $N$ spins 1/2 as

$$\hat{a}_n = \hat{a}_n(0) + \alpha_n + (\boldsymbol{u}_n + i\boldsymbol{v}_n) \cdot \hat{\boldsymbol{S}}. \tag{7}$$

Eq. (7) enables to present in a compact form some of the main achievements of our theory. This equation directly links the initial state of the atoms $\hat{\boldsymbol{S}}$ to the output state of the photons $\hat{a}_n$, which is thus fully characterized.

From Eq. (7) we can directly understand under what conditions HHG does or does not follow the conventional scenario of classical emission (that is, of harmonics in a multimode coherent state). Specifically, if the atoms are in a state with a well-defined large-$N$ classical limit, then $\hat{\boldsymbol{S}} \approx \langle \hat{\boldsymbol{S}} \rangle$ and every mode $\hat{a}_n$ is approximately a coherent state of parameter $\alpha_n + (\boldsymbol{u}_n + i\boldsymbol{v}_n) \cdot \langle \hat{\boldsymbol{S}} \rangle$. More precisely, when the atoms are all in the same single particle state $|\boldsymbol{s}\rangle$ with $\hat{\boldsymbol{\sigma}}_j|\boldsymbol{s}\rangle_j = \boldsymbol{s}|\boldsymbol{s}\rangle_j$, then $\langle \hat{\boldsymbol{S}} \rangle = N\boldsymbol{s}$ and the combined output state of multiple harmonics is a multimode coherent state $|\psi_{\text{harmonics}}\rangle \approx \prod_n |\alpha_n + N(\boldsymbol{u}_n + i\boldsymbol{v}_n) \cdot \boldsymbol{s}\rangle$. This is the case for conventional (ground state) HHG, which has $\boldsymbol{s} = -\boldsymbol{e}_z$. We note in passing that these considerations allow to maximize the yield of the $n$-th harmonic by just finding the orientation $\boldsymbol{s}$ that maximizes $|\alpha_n + N(\boldsymbol{u}_n + i\boldsymbol{v}_n) \cdot \boldsymbol{s}|$, and by preparing the atoms to such orientation with a simple coherent (rotation) pulse before the strong drive. However, if the system is in a quantum correlated many-body state lacking a clear classical limit, the replacement $\hat{\boldsymbol{S}} \to \langle \hat{\boldsymbol{S}} \rangle$ in Eq. (7)

becomes illegitimate, and $\hat{\mathbf{S}}$ should be considered with its full-fledged operatorial nature. This is the most interesting scenario, in which the emitted light can strongly deviate from a multimode coherent state.

Moreover, from a computational point of view, Eq. (7) provides a way to represent the harmonics' photon operators $\hat{a}_n$ as sparse $(N+1) \times (N+1)$-dimensional matrices. Indeed, the collective spin operators $\hat{\mathbf{S}} = (\hat{S}_x, \hat{S}_y, \hat{S}_z)$ have a standard and well-known matrix representation, and the whole complexity of the theory lies now in the terms $\alpha_n$, $\boldsymbol{u}_n$, and $\boldsymbol{v}_n$, that can be found by solving the HHG dynamics in Eq. (4) (accounting for the full atomic spectrum including its continuum) and that can be conveniently computed once and for all, in the sense that they do not change when changing the number of atoms nor their initial condition. Instead, the terms $\alpha_n$, $\boldsymbol{u}_n$, and $\boldsymbol{v}_n$ depend on the specific structure of the atoms and on the drive. As a limit case, we note that $|\alpha_n|$, $|\boldsymbol{u}_n|$, and $|\boldsymbol{v}_n|$ vanish for vanishing drive intensity, in which case the created photonic states $\hat{a}_n$ in Eq. (7) are simply the vacuum state (no emission).

The theory above allows in particular to compute normally ordered moments $\left\langle \left(\hat{a}_n^\dagger\right)^m (\hat{a}_n)^l \right\rangle = \text{Tr}\left(\hat{\rho}(0)\left(\hat{a}_n^\dagger\right)^m (\hat{a}_n)^l\right)$, for which the vacuum term $\hat{a}_n(0)$ in Eq. (7) can be omitted. Containing the full information on the emission, these moments can be used to reconstruct, for each harmonic and for a given atomic initial condition $\hat{\rho}(0)$, the Wigner function $W(\alpha)$, the photon statistics, the normalized second-order correlation function $g^{(2)}(0) = \langle \hat{a}_n^\dagger \hat{a}_n^\dagger \hat{a}_n \hat{a}_n \rangle / \langle \hat{a}_n^\dagger \hat{a}_n \rangle^2$, and the Mandel $Q$ parameter $Q = \langle \hat{a}_n^\dagger \hat{a}_n \rangle (g^{(2)}(0) - 1)$ (Supplementary Section S5). Indeed, our theory provides access to much more than that, namely, it allows to compute any multi-mode normally-ordered moment $\left\langle \left(\hat{a}_{n_1}^\dagger\right)^{m_1} \left(\hat{a}_{n_2}^\dagger\right)^{m_2} \ldots \left(\hat{a}_{n_1}\right)^{l_1} \left(\hat{a}_{n_2}\right)^{l_2} \ldots \right\rangle$, from which the full information on the joint multi-mode state of the harmonics $n_1$, $n_2$, ..., including on the entanglement among them, can in principle be reconstructed. As an example, we use the two-mode moments to compute the joint photon number statistics of two modes, and associated Parson correlation coefficient and mutual information. Finally, to help visualize the atomic state at times $t \leq 0$, we also compute the Wigner function of the atomic system on the Bloch sphere (whose axes can be thought of as corresponding to $\hat{S}_x$, $\hat{S}_y$, and $\hat{S}_z$) [37].

As for $\hat{H}_{\text{atom}}^i$ in Eq. (1), in the following we follow convention in strong field physics and model each emitter as a single electron in a one-dimensional trapping potential [22,35,36,38].

Specifically, we consider a single-particle Hamiltonian $\hat{H}_{\text{atom}}^i = \frac{\hat{p}_i^2}{2} + V(\hat{x}_i)$, with $\hat{x}_i$ and $\hat{p}_i$ the conjugate position and momentum operators of $i$-th atom, respectively, $V(x) = -\frac{1}{\sqrt{x^2+a^2}} + V_{ab}$ a softened Coulomb potential, $V_{ab}$ an imaginary potential accounting for absorbing boundaries to avoid unphysical reflections of the electrons [39]. The parameter $a$ is set to 0.816 to match the ionization potential of Ne, $I_p = 0.792$ [39].

## 2 Emitted radiation from strongly driven many-body systems

We now use our formalism to numerically investigate the properties of the radiation emitted from a many-body atomic system. We will discuss quantum features of HHG that deviate from the established expectation – that the emitted light is coherent [14,24,34] – in a way that depends on the correlations among the atoms. We consider a strong monochromatic coherent drive $E_c(t)$ at frequency $\omega_d$ and with trapezoidal pulse shape (the amplitude increases linearly to its maximum $E_0$ during the first quarter of the pulse and decreases to 0 during the last). The spectrum of the resulting emitted radiation is depicted in Figure 1b and features the distinctive traits of HHG, namely a comb of peaks at the odd harmonics extending over a plateau up to a cutoff frequency [21,22].

Beyond reproducing these well-known features of HHG, that also emerge from a classical single-atom theory [21,22], our method also captures genuinely quantum many-body ones. For instance, Figure 2 shows the photon-number statistics and Wigner function for selected harmonics and atomic initial conditions. For conventional HHG, for which all the atoms are initially in their ground state $|\Downarrow\rangle \equiv \otimes_{i=1}^{N}|g\rangle_i$ (left column in Figure 2), we confirm expectations showing an essentially coherent emission: at the odd harmonics, the correlation function $g^{(2)}$, the Wigner function, and the photon statistics are almost perfectly unitary, Gaussian, and Poissonian, respectively. In fact, the classical character of the emission persists as long as the atoms are initially in an uncorrelated product state, e.g., $|\Rightarrow\rangle \equiv \otimes_{i=1}^{N} \frac{|g\rangle_i + |e\rangle_i}{\sqrt{2}}$, which can be easily obtained from $|\Downarrow\rangle$ with a coherent $\pi/2$ pulse (central column in Figure 2). These results can be understood from Eq. (7) upon replacing $\hat{S}$ with its classical limit, namely $-N\boldsymbol{e}_z$ for $|\Downarrow\rangle$ and $N\boldsymbol{e}_x$ for $|\Rightarrow\rangle$. The emission is in this case classical, consisting of a cross-product of coherent states $|\psi_{\text{harmonics}}\rangle \approx$ $\prod_n |\alpha_n - N(\boldsymbol{u}_{n,3} + i\boldsymbol{v}_{n,3})\rangle$ and $|\psi_{\text{harmonics}}\rangle \approx \prod_n |\alpha_n + N(\boldsymbol{u}_{n,1} + i\boldsymbol{v}_{n,1})\rangle$, respectively. In this case of emission from uncorrelated atoms, any deviation of the output light from a coherent state is due to the error in the replacement $\hat{\boldsymbol{S}} \to \langle \hat{\boldsymbol{S}} \rangle$, which is however very small in the classical limit of large $N$. We can quantify the deviation from a coherent state using $g^{(2)}$, which strays from the coherent-state value 1 by $\sim 1/N$.

The situation changes drastically when many-body correlations are imprinted in the atomic state at the moment of interaction with the drive field ($t = 0$), for which the atomic state, and therefore the emission, become highly nonclassical. To test this idea, in the right column of Figure 2 we consider the system initially in the Dicke-like state $|N/2\rangle$, which is the symmetric superposition of states with half of the atoms in the ground state $|g\rangle$ and the other half in the first excited state $|e\rangle$ (within the bosonic language above, such a state is denoted $|N/2, N/2, 0,0,0, \dots\rangle$). This state is strongly correlated and its atomic Wigner function on the Bloch sphere appears as a ring with weak fringes embracing the equator. For comparison, such a Wigner function is fundamentally different from that of the uncorrelated states $|\Downarrow\rangle$ and $|\Rightarrow\rangle$, which appear as two blobs around the South and East poles, respectively. The difference in the atomic state gets mirrored onto the emission: while the radiation from $|\Downarrow\rangle$ and $|\Rightarrow\rangle$ is classical, in the sense of close to coherent, that from $|N/2\rangle$ is very much not so, its photon statistics and Wigner functions being strongly non-Poissonian and non-Gaussian, respectively.

The difference in photon statistics can be quantified by the normalized second-order correlation function $g^{(2)}$, that, for instance, for the 21st harmonics takes values 1.00002 and 1.71 for initial atomic states $|\Downarrow\rangle$ and $|N/2\rangle$, respectively. Indeed, in the latter case, the correlations initially imprinted in the atoms can result in more strikingly quantum features, such as ring-shaped Wigner functions and double-peaked photon distributions, which appear to interpolate between a thermal and a coherent state.

Intuition into the shape of the photonic Wigner function is provided by Eq. (7). Therein, the term $(\boldsymbol{u}_n + i\boldsymbol{v}_n) \cdot \hat{\boldsymbol{S}}$ acts as a projection from the three-dimensional space of the atomic Bloch sphere onto the plane individuated by $\boldsymbol{u}_n$ and $\boldsymbol{v}_n$, which becomes the plane of the two quadratures of $a_n$. The constant $\alpha_n$ corresponds to a coherent shift, and the term $\hat{a}_n(0)$ adds vacuum fluctuations with a blurring effect. In this sense, the nontrivial atomic state on the Bloch sphere gets mirrored onto a nontrivial output photonic state.

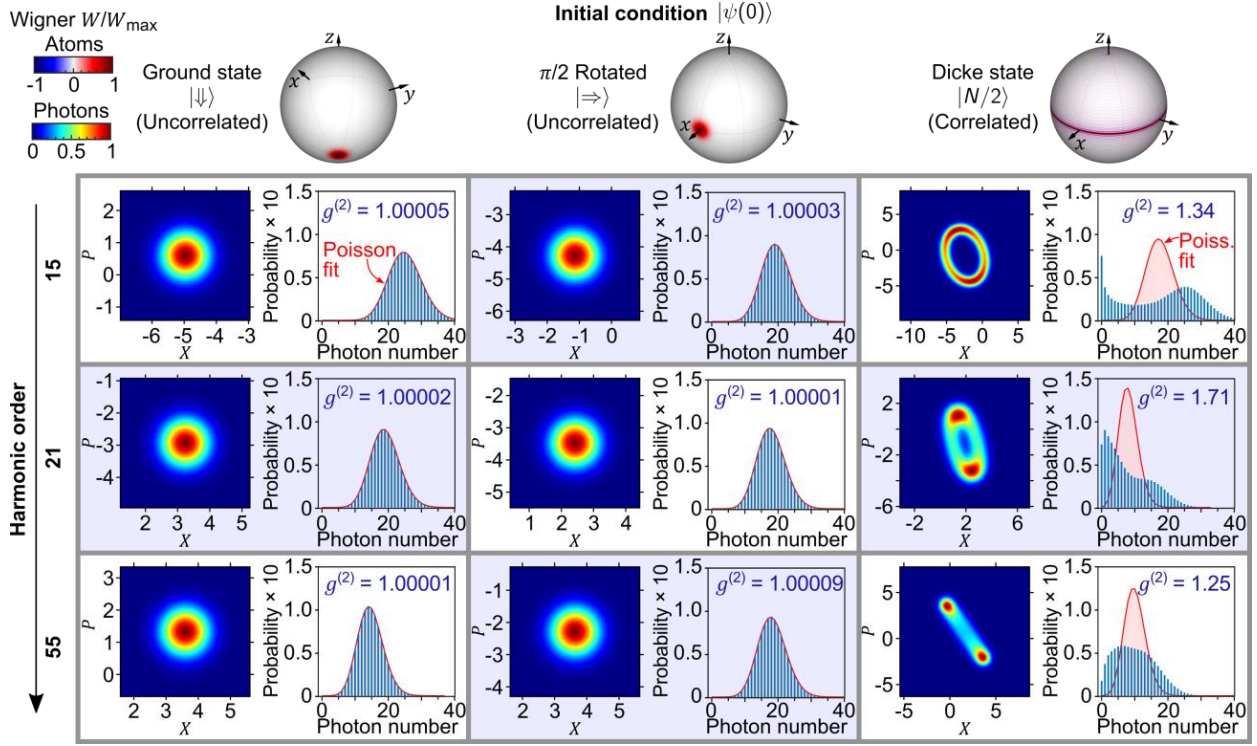

**Figure 2. Many-body high-harmonic generation (HHG).** We investigate the emission for some representative odd harmonics ($n = 15$, $21$, and $55$ in the top, middle, and bottom rows, respectively) and from selected atomic initial conditions ($|\Downarrow\rangle$, $|\Rightarrow\rangle$, $|N/2\rangle$ in the left, center, and right columns, respectively). To help visualize the atomic initial conditions, we plot their Wigner function on the Bloch sphere. (**Left**) The HHG emission from atoms initially in the ground state $|\Downarrow\rangle$ is essentially coherent and, thus, classical. The Wigner function of the input atomic state is Gaussian (around the South pole on the Bloch sphere), like the Wigner function of the output HHG emission. The photon number statistics is excellently fitted by a Poisson distribution (in red), and the normalized second-order correlation function $g^{(2)}$ (in blue) is close to 1 (deviations are of order $1/N$). (**Center**) The classical character of the emission holds for other uncorrelated initial conditions, such as the state $|\Rightarrow\rangle$, obtained from $|\Downarrow\rangle$ with a $\pi/2$ coherent pulse. (**Right**) In striking contrast, strong correlations in the initial condition, e.g., for the state $|N/2\rangle$, whose atomic Wigner function embraces the Bloch sphere like a belt around the equator, result in highly nonclassical light. Indeed, the Wigner distribution and photon statistics at the odd harmonics are strongly non-Gaussian and super-Poissonian, respectively. For instance, in the case of the 15$^{\text{th}}$ harmonic we observe a doubly-peaked photon statistics. The simulations in these panels assume $\hbar\omega_d = 1.55$ eV, $E_0 = 60$ GV/m, $N = 62000$, 40 cycles of the drive. The atomic Wigner functions were obtained for $N = 100$.

## 3 Engineering and controlling novel states of light

In the previous Section, simulating emission from the state $|N/2\rangle$ helped emphasize the role of atomic correlations in creating new states of light, showing the concept of nonclassical light emission from nonclassical states of the emitters [40,41]. However, the Dicke-like state $|N/2\rangle$ might be challenging to realize in experiments, and thus we propose two different schemes that use coherent control [42] (starting from atoms initially in their ground state) to induce atomic correlations that resemble the ones of the $|N/2\rangle$ state, resulting in nonclassical light emission akin to that from $|N/2\rangle$ (Figures 3 and 4). In each scheme, we find the same concept: that the strongly driven many-body systems can generate bright and strongly non-coherent radiation. The schemes that we propose consist of three main steps: (i) A short preparation pulse brings the atoms into a coherent superposition of excited and ground states; (ii) Correlations between atoms build up under the system's own dynamics throughout an hold time $t_h$; (iii) The main driving pulse is applied, resulting in HHG emission with unconventional properties. For concreteness, we will consider the scenario in which correlations result from interatomic interactions, and one in which they are induced through superradiance.

Let us begin by considering the scenario in which atomic correlations are induced through superradiance [29,30] (Figure 3). Superradiance is the phenomenon whereby the spontaneous emission from an ensemble of excited emitters can be much stronger than one would expect if they emitted independently. With a long history, superradiance has played an important role in optics and quantum mechanics, for instance with applications in quantum metrology [43], as well as in relativity and astrophysics [44]. Crucially, superradiance creates quantum correlations among the atoms, induced by their joint interaction with the surrounding electromagnetic field. Here, we exploit this feature of superradiance to controllably prepare the atoms in a correlated state, resulting in nonclassical HHG upon shining the system with a strong drive. The specific protocol we consider is illustrated in Figure 3a, and unfolds as follows. A coherent $\pi$-pulse brings the atoms from $|\Downarrow\rangle$ to $|\Uparrow\rangle \equiv \otimes_{i=1}^{N} |e\rangle_i$ at time $t = -t_h$. From $|\Uparrow\rangle$, the system tends to decay back to $|\Downarrow\rangle$ via superradiance, which involves the build up of correlations between atoms. The superradiance process is however intermitted by the beginning of the HHG pulse after a time $t_h$, i.e., at time $t = 0$. The parameter $t_h$ thus offers a handle to control the initial atomic condition $\hat{\rho}(0)$ of the HHG emission. More explicitly, we find $\hat{\rho}(0)$ by numerically integrating from $t = -t_h$ to $t = 0$ the following collective Lindblad Master equation describing superradiance [30,45,46]

$$\frac{d\hat{\rho}}{dt} = \gamma \left( \hat{S}^- \hat{\rho} \hat{S}^+ - \frac{1}{2} \{\hat{S}^+ \hat{S}^-, \hat{\rho}\} \right), \tag{8}$$

with $\gamma$ the emission rate and starting from $\hat{\rho}(-t_h) = |\Uparrow\rangle\langle\Uparrow|$. Eq. (8), whose solution is facilitated by the diagonal nature of $\hat{\rho}(t)$, captures the distinctive bell-shaped time profile of the superradiance emission intensity, with maximum emission after a time $t_m \approx \frac{\log N}{4\gamma N}$ [30], as well as the buildup of atomic correlations. The state of the system through the superradiance process can be visualized in term of the atomic Wigner function on the Bloch sphere, see Figure 3a. The atomic Wigner function, initially concentrated around the North pole for $\hat{\rho}(-t_h) = |\Uparrow\rangle\langle\Uparrow|$, "cascades" around the Bloch sphere during superradiance. Halfway through the decay, the atomic Wigner wraps the sphere in a way that is reminiscent of the Wigner function of the Dicke state $|N/2\rangle$ in Figure 2, although with a much wider broadening along the z axis. This perspective suggests that we should be able to obtain nonclassical emission akin to that from $|N/2\rangle$.

The second order correlation function $g^{(2)}(0)$ and Mandel $Q$ parameter of the emission resulting from a strong drive are shown in Figure 3b for a few selected harmonics. A classical emission ($g^{(2)}(0) = 1$ and $Q = 0$) is obtained for both $t_h = 0$, for which superradiance does not have the time to start and $\hat{\rho}(0) = |\Uparrow\rangle\langle\Uparrow|$, and for $t_h \gg t_m$, for which superradiance makes all atoms decay and $\hat{\rho}(0) \approx |\Downarrow\rangle\langle\Downarrow|$. In between these two limit cases, $\hat{\rho}(0)$ is nontrivial and accounts for correlations among the atoms, which directly translates onto a nonclassical emission with harmonics strongly deviating from a coherent state ($g^{(2)}(0) > 1$ and $Q > 0$). The full portrait of the harmonics is given in Figure 3c in terms of photonic Wigner functions. These are approximately Gaussian for $t_h = 0$ and $t_h \gg t_m$, while acquire a richer structure for intermediate $t_h$. In particular, for $t_h = t_m$, that is, shining the strong drive pulse onto the system when its superradiance intensity is at its maximum, we find as hoped photonic Wigner functions reminiscent of those obtained for $|N/2\rangle$ in Figure 2, although more blurred.

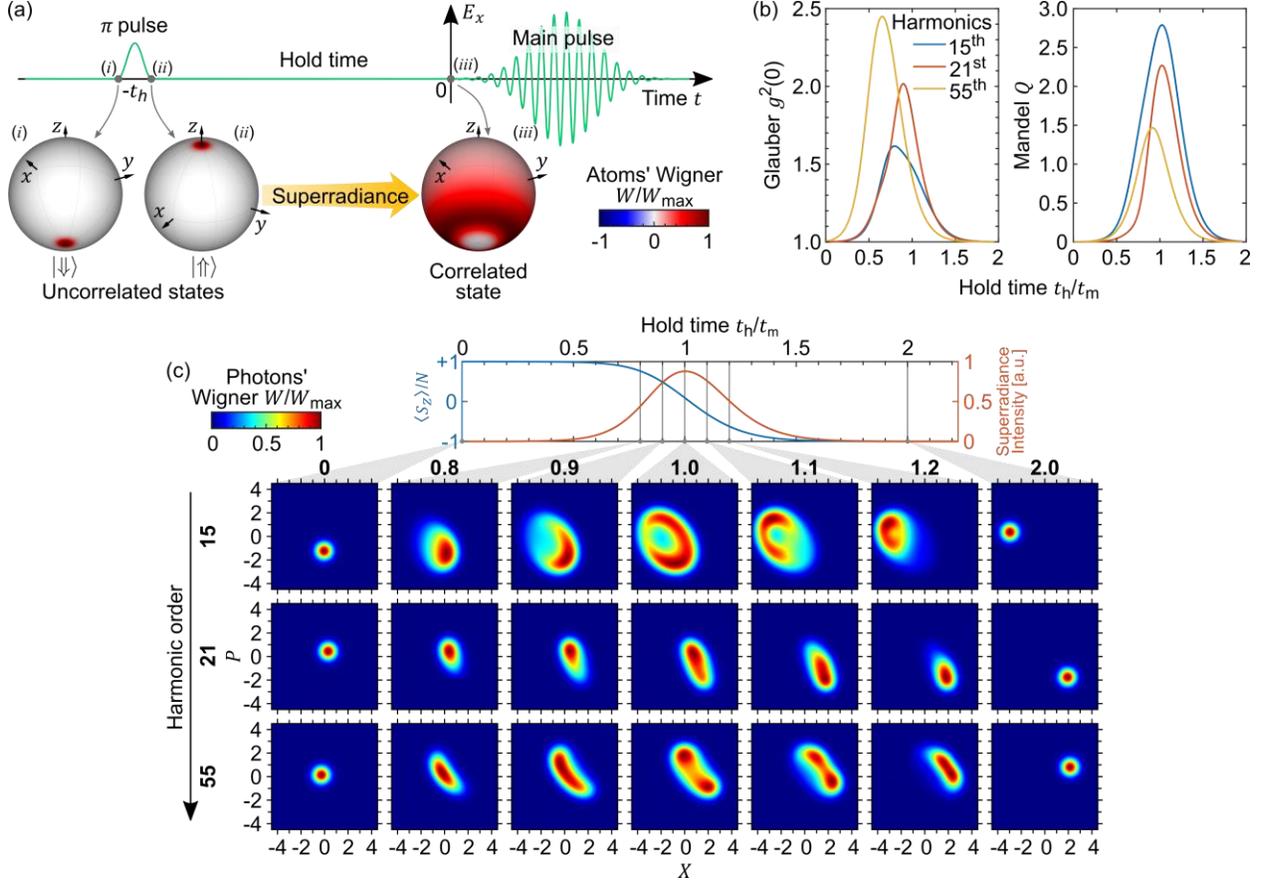

**Figure 3. Quantum light from strongly driven superradiant atoms.** Superradiance is exploited as a mean to induce correlations in the atomic initial condition $\hat{\rho}(0)$, leading to unconventional HHG. (**a**) Schematic representation of the adopted three-step protocol (i) the atoms are excited from $|\Downarrow\rangle$ to $|\Uparrow\rangle$ with a $\pi$-pulse, (ii) decay through superradiance over a hold time $t_h$, (iii) and are shined with a strong drive pulse leading to HHG. (**b**) The correlation function $g^{(2)}(0)$ and Mandel $Q$ parameter of the emitted harmonics depend on the amount of atomic correlations at $t = 0$, that is, on the hold time $t_h$. (**c**) The Wigner functions of the harmonics are approximately Gaussian for $t_h = 0$ (for which $\hat{\rho}(0) = |\Uparrow\rangle\langle\Uparrow|$) and $t_h \gg t_m$ (for which $\hat{\rho}(0) \approx |\Downarrow\rangle\langle\Downarrow|$), whereas they develop non-coherent features for intermediate hold times $t_h \sim t_m$. In particular, for $t_h = t_m$ the emission is highly nonclassical and looks reminiscent of that from the Dicke state $|N/2\rangle$. On top, the time profiles of atomic magnetization and emission intensity are reported. These simulations considered $\omega_d = 1.55$ eV, $E_0 = 60$ GV/m, $\gamma N = 0.1$, $N = 37000$, and 40 cycles of the drive. The atomic Wigner functions on the Bloch spheres in (**a**) were obtained for $N = 100$ and $t_h = 1.3\, t_m$.

Second, we exemplify the concept of HHG from a many-body system correlated through interactions by considering the case in which, prior to the pulse (i.e., at $t < 0$) the atomic system is described by the one-axis twisting Hamiltonian [31]

$$\hat{H} = \frac{\omega_0}{2}\hat{S}_z + \frac{\omega_J}{N}\hat{S}_z^2. \tag{9}$$

In Eq. (9), the first term is nothing but the original $\hat{H}_{\text{atom}}^i$, just restricted to the two lowest single-particle levels $|g\rangle$ and $|e\rangle$, with $\omega_0 = 0.49$ the energy difference between them. The second term, of strength $\omega_J$ and normalized by $N$ to guarantee extensivity, accounts instead for a collective interaction. This can for instance occur in trapped cold atoms in optical cavities [47,48], and is standard in the context of spin squeezing [31]. But most importantly, from a theory perspective, the Hamiltonian (9) preserves the atoms' permutational symmetry, allowing the theoretical framework above to be readily applied.

The protocol we consider, illustrated in Figure 4a, closely follows that introduced by Kitagawa and Ueda in their seminal work for spin squeezing [31]: a $\pi/2$ pulse brings the atoms in the state $|\Rightarrow\rangle$, from which correlations develop under the action of the twisting term $\hat{S}_z^2$ in Eq. (9). The action of the latter can be effectively visualized in terms of the atomic Wigner function on the Bloch sphere, that gets progressively more and more sheared, wrapping around the Bloch sphere to embrace its equator. This perspective suggests how, for long enough hold times $t_h$, the state of the atoms becomes akin to that of $|N/2\rangle$, whose Wigner function was indeed a ring encircling the equator (see Figure 2). By tuning the hold time $t_h$ separating the two pulses, we can therefore interpolate from a regime of classical emission from an uncorrelated atomic state $|\Rightarrow\rangle$ to one of strongly nonclassical emission form a strongly correlated atomic state akin to $|N/2\rangle$. In Figure 4b, this is shown in terms of second order correlation function $g^{(2)}(0)$ and Mandel $Q$ parameter, taking for $t_h = 0$ values 1 and 0, respectively and corresponding to coherent emission, and larger values for $t_h > 0$. In Figure 4c we instead show the Wigner function of the emitted light, showing how starting from Gaussian for $t_h = 0$, it gets deformed for $t_h > 0$, becoming for long enough $t_h$ completely analogue to that resulting from atoms initially in $|N/2\rangle$, which we report as a reference.

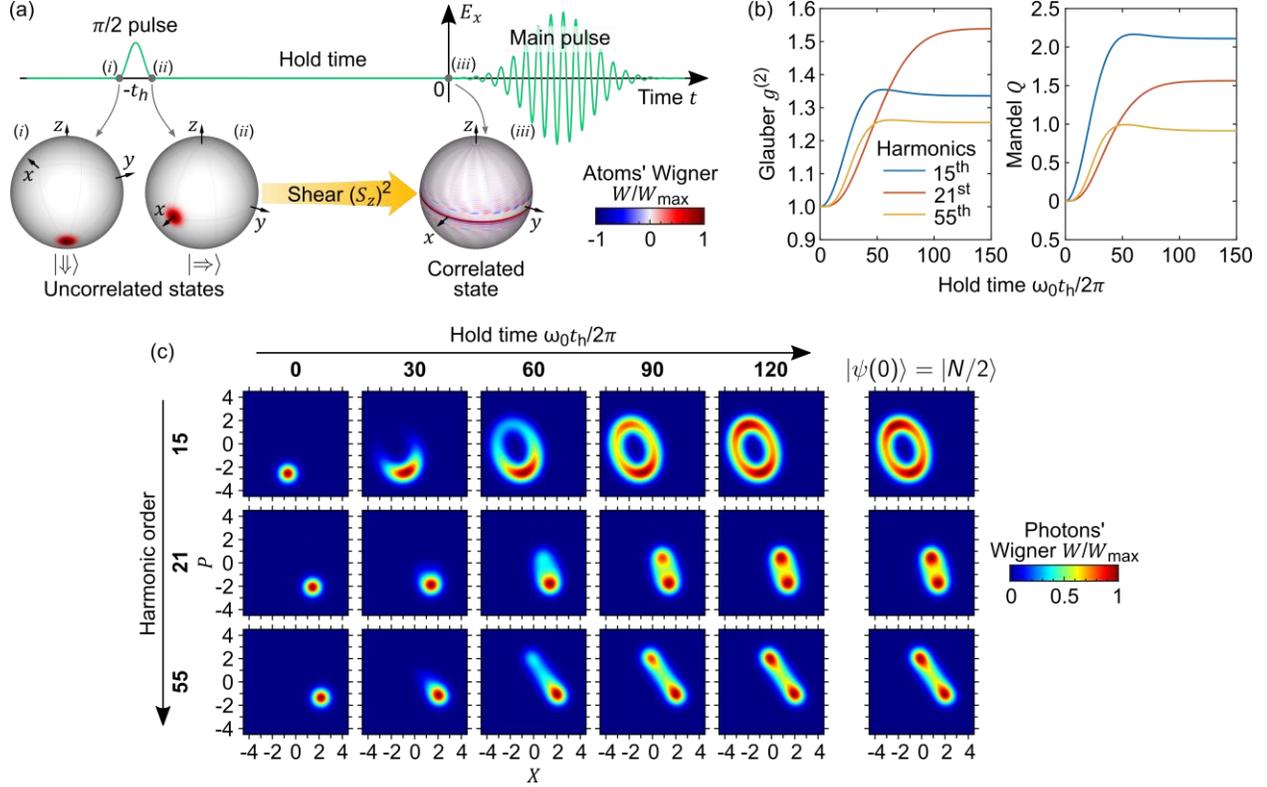

**Figure 4. Quantum light from strongly driven interacting atoms.** The physics and structure of this figure are in analogy to Figure 3, but exploiting inter-atomic interactions, rather than superradiance, to induce the atomic correlations underpinning nonclassical HHG. **(a)** Specifically, we consider that: (i) a preparation $\pi/2$-pulse is applied to atoms initially in their ground state, (ii) atomic correlations build up via interatomic interactions throughout a hold time $t_h$, (iii) the system is illuminated with an intense drive laser pulse resulting in HHG. **(b)** The correlation function $g^{(2)}(0)$ and Mandel $Q$ parameter of the emitted light depend on the amount of correlations in the atomic system at the time of the strong pulse, that is, on the hold time $t_h$, with $t_h = 0$ the limit of classical coherent emission from uncorrelated atoms. **(c)** The properties of the emission are more comprehensively illustrated by the Wigner distributions of the harmonics of interest. Starting from Gaussian for $t_h = 0$, these are deformed for $t_h > 0$, becoming for $t_h \omega_0/2\pi \gtrsim 120$ very close to those obtained for atoms initially in $|N/2\rangle$ (reported in the right column as a reference). Here, we considered $\omega_d = 1.55$ eV, $E_0 = 60$ GV/m, $\omega_J = 2.7$ eV, $N = 37000$, and 40 cycles of the drive. The atomic Wigner functions on the Bloch spheres in (a) were obtained for $N = 100$ and $\omega_J t_h = 50$.

We finally investigate the correlations between different harmonics. The qualitative finding is that, while the harmonics are in a separable state in conventional ground state HHG, they become correlated when the atoms are prepared in a correlated state. To facilitate this finding we focus for concreteness on the second scheme considered above, that of interaction-induced correlations, and compute the joint photon statistics between two harmonics $n$ and $m$, that is, the probability $p(k_n, k_m)$ to observe $k_n$ photons in the mode $n$, and $k_m$ photons in the mode $m$ (Figure 5a). By changing $t_h$ we can again interpolate between an uncorrelated atomic state for $t_h = 0$, and a correlated atomic state akin to $|N/2\rangle$ for large $t_h$. To quantify the degree of interdependence between the two harmonics, from the joint photon number statistics $p(k_n, k_m)$ we compute the Parson correlation coefficient $c = \frac{\text{cov}(k_n, k_m)}{\sigma_{k_n} \sigma_{k_m}}$, with $\text{cov}(k_n, k_m)$ the covariance of the photon numbers and $\sigma_{k_n}$ and $\sigma_{k_m}$ their standard deviations, and the mutual information $I_{mn} = \sum_{k_n, k_m} p(k_n, k_m) \log \frac{p(k_n, k_m)}{p(k_n) p(k_m)}$. In Figure 5b we show these diagnostics to vanish for $t_h = 0$, meaning that the harmonics are uncorrelated if the atoms are. By contrast, initially correlated atoms ($t_h > 0$) result in statistically dependent ($I_{mn} > 0$) harmonics. The correlation between two harmonics can be both positive ($c > 0$, as for the 21$^{\text{st}}$ and 55$^{\text{th}}$ harmonics) or negative ($c < 0$, as for the 15$^{\text{th}}$ and 21$^{\text{st}}$ harmonics).

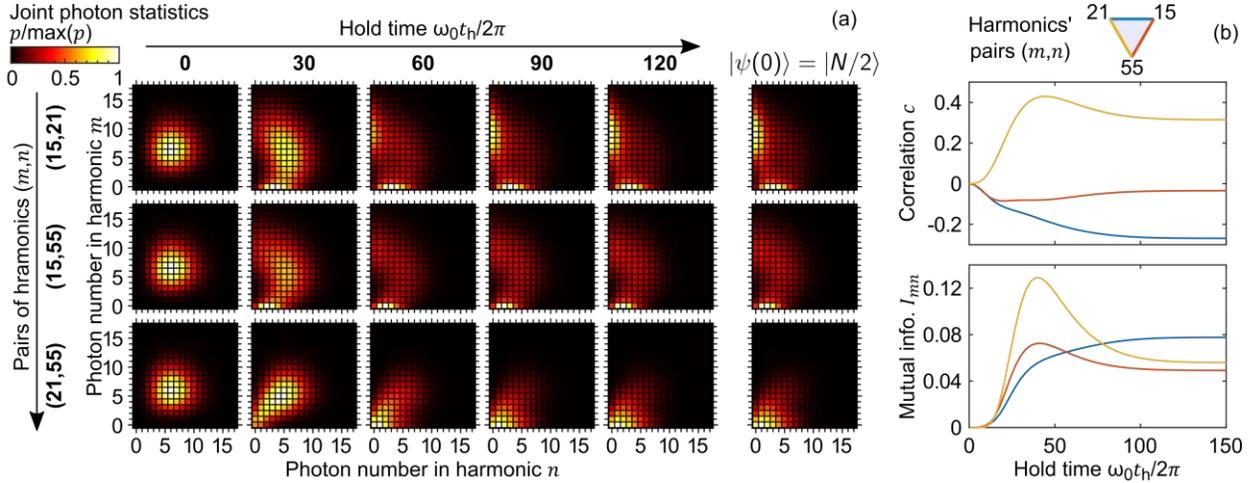

**Figure 5. Correlated states of light.** Using the same protocol as in Figure 4, we show that preparing the atoms in a correlated state results in strong correlations between different output harmonics. (**a**) The joint photon statistics for selected pairs of harmonics (one pair per row) becomes increasingly nontrivial for larger hold times $t_h$ (one time per column), that is, for stronger initial atomic correlations. For $t_h \omega_0 / 2\pi \gtrsim 120$, the joint photon statistics becomes very close to that obtained from the state $|N/2\rangle$, reported in the last column as a reference. (**b**) The Parson correlation coefficient $c$ and the mutual information $I_{mn}$ are witnesses of the entanglement between harmonics pairs, vanish for $t_h = 0$ (that is, uncorrelated atomic initial condition), and are finite otherwise. Here, we used the same parameters as in Figure 4.

**Path towards experimental realization**

Many of the quantum-optical properties we predict for HHG should be within the reach of state-of-the-art experiments. The photon statistics, Glauber correlation coefficient $g^{(2)}$, and Mandel parameter $Q$ of the harmonics of interest, can all be computed from photon counting measurements averaged over repeated experiments. Such measures should clearly distinguish emission by atoms imprinted with strong correlations relative to emission of classical light (e.g., $g^{(2)} > 2$ in Figure 3 vs $g^{(2)} = 1$ for a coherent state). Indeed, for uncorrelated atoms in conventional HHG, the deviation of the emitted harmonics from a (classical) perfectly coherent multimode state are small, as shown in Figure 2 and in [14,24]. The joint photon statistics, correlation coefficient, and mutual information between two harmonics (Figure 5), could also be extracted by photon counting detectors. Finally, the Wigner function of the emitted harmonics could be reconstructed via homodyne detection [49–51]. The lower harmonics could be characterized in this way by established conventional means for homodyne detection in the visible range. Homodyne schemes for the higher harmonics (UV, XUV, or soft-X-ray frequencies) will require new developments, which are gradually becoming feasible (e.g., interferometers and sub-cycle delay lines in the XUV [52–54]).

In our one-dimensional atomic model, we set $a = 0.816$ to best mimic the spectrum of Ne. The choice of potential can be readily revised to model other emitting systems. In fact, more realistic simulations can be obtained by considering three-dimensional atomic Hamiltonians for $\widehat{H}_{\text{atom}}^i$, or directly replacing the particle energies $\boldsymbol{W}$ and dipole moments $\boldsymbol{D}$ in Eq. (4) by values extracted from a density functional theory (DFT) simulation. This choice does not change the structure of the theory itself, and thus the conclusions are independent of the details of the model. The form and interpretation of Eq. (7) remain unchanged (only the values of $\alpha_n$, $\boldsymbol{u}_n$, and $\boldsymbol{v}_n$ would change), meaning that the qualitative results we found carry over to more realistic scenarios. Indeed, this establishes the generality of the concept that we put forward, that of nonclassical emission from correlated strongly driven many-body systems.

A core assumption in our theoretical derivation is that of indistinguishable atoms. This Dicke-like assumption makes the notoriously hard many-body problem tractable. The main findings of the theory are conveyed by Eq. (7), which shows the transfer of the nonclassical states of the input emitters to the nonclassical states of the output light. The assumption of indistinguishable atoms is legit, for instance, when large collections of atoms occupy a small volume as compared to the involved radiation wavelengths. To have a concrete estimate of the number of atoms $N$ fulfilling this criterion, consider a drive wavelength of $\lambda = 800$ nm creating the $n = 15$th harmonic, with wavelength $\lambda_n = \lambda/n = 53$ nm. The volume in which atoms can be considered indistinguishable is $V \sim (\lambda_n/2)^3$. For an ideal gas, the number of particles within such a volume reads $N = \frac{pV}{RT} N_a$, with $p$, $R$, $T$, and $N_a$ the presusre, gas constant, Avogadro number, and temperature, respectively. At room temperature and atmospheric pressure, we get $N \sim 500$, and larger $N$ values similar or beyond those simulated in in Figures 3, 4, and 5 could be obtained for example just increasing the pressure and/or lowering the temperature [55]. The number $N$ can be even greater within the emerging field of solid-state HHG, which is now widely explored experimentally (e.g., [56–58]). Given that the atomic spacing between atoms in solids is usually of a few Angstroms, one finds $N \sim 10^5 - 10^6$ for the same volume above. Interestingly, it might be possible to arrange the atoms over a larger volume (extending over many wavelengths) such that they can be effectively considered indistinguishable, e.g., thanks to phase matching in atomic ensembles [45,46].

In any case, the general concept of nonclassical emission in HHG from nonclassical many-body states of emitters is likely much broader than the indistinguishability assumption that we used here to show it. Indeed, numerous many-body phenomena have traditionally been first studied within a similar assumption, before being extended far beyond it. This idea applies to countless phenomena in and out of equilibrium, including those leveraged by the preparation schemes adopted in Figures 3, 4, and 5 here, namely superradiance [29,30] and spin squeezing [31]. We expect the theory of strongly driven many-body systems to follow a similar path.

The approximation of indistinguishable atoms has here also been adopted to model the preparation stages, in which correlations in the atomic ensemble are induced via either superradiance or interatomic interactions. In the case of superradiance, the indistinguishability assumption is for example justified when the atoms are within a volume $V \sim (\lambda_0/2)^3$ [30], where $\lambda_0 = 92$ nm is the wavelength associated to the transition $|e\rangle \to |g\rangle$ and that is anyway a less stringent condition than that analysed above with the wavelength $\lambda_n$ of the harmonic $n = 15$, given that $\lambda_0 > \lambda_n$. In the case of interaction-induced correlations, relevant systems include strongly interacting gases of atoms or molecules, whose interactions can be enhanced by trapping them in optical cavities. Such systems can realize the infinite-range (all-to-all) $\hat{S}_z^2$ interactions (shear-force terms) discussed in the previous Section [47,59]. Moreover, finite-range spin-spin interactions, which are ubiquitous in many hot-vapor [60] and cold-atom systems [61], can also realize spin-squeezed states [62,63] and thus should also show variants of the quantum features we proposed here. Another promising platform for exploring these ideas is Rydberg atoms [47,48], [64], offering the possibility of controllably generating correlated many-body atomic states [65,66].

Note that the collective forms of the superradiant decay and interactions in Eqs. (8) and (9), respectively, have been chosen for pure tractability reasons, and do not represent a limitation to the validity of our general concept of emission by strongly driven correlated many-body systems. In fact, any procedure that creates nonclassical states of atoms can result, in light of Eq. (7), in nonclassical output harmonics.

Finally, Dicke states such as $|N/2\rangle$ could in principle be prepared via post-selection in superradiance. Specifically, if one could measure the number $n_{super}$ of photons spontaneously emitted during superradiance [described by Eq. (8)], the atoms would then collapse in the state $|N - n_{super}\rangle$ consisting of a symmetric superposition of $n_{super}$ atoms in the ground state $|g\rangle$ and $N - n_{super}$ atoms in the first excited state $|e\rangle$. For $n_{super} \approx N/2$, this initial condition would lead to emission similar to that obtained for $|N/2\rangle$ in Figure 2.

## Discussion and outlook

Our work proposes strongly driven many-body systems to realize many-photon states of light with controllable quantum features and over a broad spectral range up to X-ray frequencies. At the core of this concept is the idea that nonclassical states of the emitters get reflected onto nonclassical states of the emission. For example, Eq. (7) shows how the Wigner function of the atoms (on the Bloch sphere) gets effectively projected onto that of the emitted harmonics (in the plane of the two quadratures of light), see, e.g., Figure 2. The paradigm of classical coherent emission in HHG, well established for conventional HHG from atoms initially in their single-particle ground states, can now be overcome: if the atoms are strongly correlated, the output harmonics exhibit strongly noncoherent features, such as doubly peaked photon statistics and intra-harmonic correlations. The challenge of engineering strongly nonclassical states of light is therefore shifted to the comparably simpler one of preparing strongly correlated atomic states. We considered just but two examples, in which the atomic correlations are dynamically generated under the system own (undriven) dynamics during a preparation stage, via either superradiance or interatomic interactions acting throughout a hold time $t_h$ between an excitation and the HHG pulses. Varying $t_h$, one can control the amount of correlations in the atoms, and therefore the deviation of the harmonics from a multimode coherent state.

The ideas presented here have several potential applications. We showed that HHG emission from a correlated system has super-Poissonian statistics and a broadband spectrum (e.g., $g^{(2)} > 2.0$ in Figure 3). Such bunched light could be used to enhance the efficiency of certain nonlinear processes driven by the high-harmonics. For comparison, in the optical range, it has been shown that for bright squeezed vacuum light (for which $g^{(2)} = 3$) the efficiencies of the $\chi^{(2)}$ and $\chi^{(3)}$ nonlinearities increase by a factor $\sim 3$ and $\sim 14$, respectively [10]. The increased efficiencies motivate using our bunched high-harmonics to drive processes that relate to $\chi^{(2)}$ and $\chi^{(3)}$, such as second and third harmonic generation, which could lead to even higher frequencies, producing super-Poissonian X-ray light.

The correlations between different harmonics offer new prospects for enhancing the signal-to-noise ratio in quantum imaging [67]. Such correlations have not been predicted by previous theories of HHG. The correlations between widely different harmonics could be particularly relevant for biological samples, which are very sensitive to ionizing radiation and specifically X-ray light in the water-transparency window that can be reached using HHG [68]. For instance, we envision using the correlations between harmonics to acquire an image by measuring photons of a different frequency than the one interacting with the target sample.

Finally, our work suggests HHG as a probe to characterize many-body states of matter. For instance, by measuring the Wigner function of the output harmonics [49–51], it should be possible via Eq. (7) to reconstruct the atomic Wigner function, thus accessing information on the many-body state of the atoms with high temporal resolution [19,20]. One of the most interesting implementations of our general concept would be to try to image the attosecond dynamics of vortices in superfluids and superconductors [69], Bose-Einstein condensates [70], and other strongly correlated many-body systems. The quantum correlations in such phenomena could be transferred to the quantum state of light, and thus inferred by quantum-optical means.

The theory that we developed can be adopted or generalized in several different research directions. One question worthwhile pursuing regards further ways of controlling the atomic initial conditions and thus the emitted radiation, for instance involving more than just the two lowest single-particle levels. Beyond HHG, our theory can then be readily generalized to any strongly driven many-body system that emits radiation: further work should investigate the possibility of engineering not only the initial condition of the atoms, but the temporal profile of the driving pulse as well. An ambitious generalization of the theory could investigate the effects of the spatial distribution of the atoms, e.g., in solids, going beyond the approximation of indistinguishable atoms. Looking forward, our work contributes to the ambitious goal of bringing together quantum optics and attoscience, suggesting a new path towards the realization of fully tuneable sources of intense quantum light in new spectral ranges.

## Acknowledgments

The authors wish to acknowledge insightful discussion on related topics with R. Bekenstein, O. Cohen, E. G. Dalla Torre, M. Even Tzur, M. Faran, R. Ruimy, E. Shahmoon, and J. Sloan. The authors are especially thankful to D. Malz and M. Kruger for discussions and for comments on the manuscript. A. P. acknowledges support from the Royal Society, support from the AFOSR MURI program (Grant No. FA9550-21-1-0069), and hospitality at TUM.

## Author contributions

All the authors gave critical contributions to this work.

## Competing interests' statement

The authors declare no competing interests.

# Supplementary Information for

# "Light emission from strongly driven many-body systems"

Andrea Pizzi, Alexey Gorlach, Nicholas Rivera, Andreas Nunnenkamp, and Ido Kaminer

This Supplementary Information is devoted to technical derivations and complementary results and is structured as follows. We begin in Section 1 by presenting the one-dimensional single atom Hamiltonian. In Section 2, we describe the preparation protocols for inducing correlations in the atoms, based on either superradiance or spin squeezing from interatomic interactions. In Section 3, we show how permutationally invariant atomic states and operators can be described in terms of bosonic operators, in close analogy with second quantization. In Section 4, we exploit the bosonic representation of the atomic degrees of freedom to build a theory of many-body HHG in the Heisenberg picture. The ultimate achievement of such a section is the computation of the normally ordered moments $\left\langle \left(\hat{a}_n^\dagger\right)^m (\hat{a}_n)^l \right\rangle$ for each $m$, $l$, and harmonic order $n$. In Section 5, we show how the information on the normally ordered moments can be used to extract the Wigner function and the photon statistics of the emitted harmonics. In Section 6 with a hands-on summary outlining how to practically implement our theory in a numerical simulation. In Section 7 we present a few complementary results. In Section 8 we present an alternative phase space approach that, while not being exact, allows an efficient and qualitatively correct description of the output emission.

*Note*: Throughout this Supplementary Information we adopt atomic units unless otherwise specified. Moreover, we use the notations $|⇑⟩$ and $|↑↑ \cdots ↑⟩$ interchangeably (and similarly for $|⇓⟩$ and $|↓↓ \cdots ↓⟩$ and for $|⇒⟩$ and $|→→ \cdots →⟩$).

## Section S1: Single-atom Hamiltonian

We describe the atoms through a one-dimensional single-particle Hamiltonian, as standard in the context of HHG [1–3].

$$\hat{h} = \frac{\hat{p}^2}{2} + V(\hat{x}), \qquad (S1)$$

with $\hat{x}$ and $\hat{p}$ conjugate position and momentum operators. In the position basis, $\hat{x} = x$ and $\hat{p} = -i\partial_x$. We adopt a softened Coulomb potential [4,5], reading

$$V(x) = -\frac{1}{\sqrt{x^2 + a^2}} + V_{ab}(x), \qquad (S2)$$

where we set $a = 0.816$ to match the ionization potential of Ne, that is $I_p = 0.792$, and where $V_{ab}$ is an absorbing boundary potential to prevent nonphysical reflection of the electron form the grid boundaries [5],

$$V_{ab}(x) = \begin{cases} -5i \times 10^{-4}(|x| - x_0)^3 & if\ |x| > x_0 \\ 0 & if\ |x| < x_0 \end{cases}. \qquad (S3)$$

The single single-particle energy eigenstates $|m⟩$ are defined by,

$$\left(\frac{\hat{p}^2}{2} + V(\hat{x})\right)|m\rangle = w_m|m\rangle, \tag{S4}$$

with $w_m$ the eigenenergy. Note that the complex potential $V_{ab}$ introduces imaginary parts in the eigenstates of the free states, while having negligible effects on the bound states. The atomic spectrum is found numerically via standard space discretization. We consider a box $-L < x < L$, place the absorbing boundaries at $\pm x_0 = \pm 0.9L$, and discretize the space with step $dx$. The resulting spectrum is composed of $M = \frac{2L}{dx}$ levels.

The choice of the parameters $L$ and $dx$ is helped by the following heuristic considerations. (i) The ponderomotive energy of the electrons is $U_p = \frac{E_0^2}{4\omega_d^2}$. We consider driving field strengths $E_0$ such that $U_p \gtrsim I_p \approx 0.792$, that is, $E_0 \sim 2\omega_d$. (ii) The amplitude of oscillation of a free electron under the periodic drive is $x_{max} = \frac{E_0}{\omega_d^2} \sim \frac{2}{\omega_d}$: the boundaries $L$ should be larger than $x_{max}$, that is $L > \frac{2}{\omega_d}$. (iii) For free electrons in a discrete space, the discrete spectrum reads $E_k = \frac{1}{(dx)^2}\left(1 - \cos\frac{k\pi}{N_x}\right)$ for $k = 1, 2, \ldots, N_x$ and $N_x = \frac{2L}{dx} + 1$. On the one hand, the level spacing $\delta E \sim \frac{\pi}{2Ldx}$ should be, for a good representation of the continuum, smaller than the other energy involved scales $\omega_d$ and $E_0$. Thus, we set $\frac{\pi}{2Ldx} < \omega_d$, that is, $Ldx > \frac{2}{\omega_d}$. On the other hand, the largest of the levels $E_{max} = \frac{2}{(dx)^2}$ should be larger than the maximum energy of an electron under the periodic drive, that is $2U_p$, from which we ask $\frac{2}{(dx)^2} > \frac{E_0^2}{4\omega_d^2}$, that is, $dx < \sqrt{2}$. (iv) Furthermore, $dx$ should be small enough to resolve the atomic potential. More specifically, taking the distance $\sqrt{2}a$ between the two inflection points of the potential $V$ as its width, we need $dx < \sqrt{2}a$. This condition is more stringent than that from (iv), and therefore the one we retain. Putting all these conditions together, we require

$$dx < \sqrt{2}a, \quad L\,dx > \frac{2}{\omega_d}, \tag{S5}$$

Clearly, decreasing $dx$ and increasing $L$ comes at a cost, because it implies larger numbers of discrete levels ($M \approx 2L/dx$) and thus more demanding computations. All in all, we find that a good choice of parameters, for the considered $\omega_d = 0.057$ (corresponding to a wavelength of $\lambda = 800\,nm$), is $L = 150$ and $dx = 0.7$, for which the calculation involves $M = 429$ levels.

## Section S2: Bosonic representation of strongly driven atoms

In this Section, we show how a general system of identical particles in a symmetrized state can be described by a set of bosonic operators associated with the single-particle states, in the same spirit of second quantization or the Schwinger mapping for angular momentum [6].

Consider a collection of $N$ identical particles and a basis of single-particle states $\{|m\rangle\}$. Given a permutation $p = (p_1, p_2, \ldots, p_N)$, a permutation operator $\hat{P}_p$ can be defined by

$$\hat{P}_p|m_1, m_2 \ldots m_N\rangle = |m_{p_1}, m_{p_2} \ldots m_{p_N}\rangle, \tag{S6}$$

where $|m_1, m_2 \ldots m_N\rangle$ is the state with $j^{th}$ particle is in state $|m_j\rangle$. A symmetrization operator creating symmetric states invariant under any permutation is then defined as

$$\hat{S} = \frac{1}{\sqrt{N!}} \sum_p \hat{P}_p, \tag{S7}$$

where the sum runs over all the $N!$ possible permutations $p$. The symmetric (Fock) state with $n_1$ particles in the state $|1\rangle$, $n_2$ in the state $|2\rangle$,…, $n_m$ in the state $|m\rangle$ is denoted as

$$|\boldsymbol{n}\rangle = |n_1, n_2 \ldots, n_m, \ldots\rangle = \frac{1}{\sqrt{\prod_m n_m!}} \hat{S} |\underbrace{1,1,\ldots,1}_{n_1 \text{ times}}, \underbrace{2,2,\ldots,2}_{n_2 \text{ times}}, \ldots, \underbrace{m,m,\ldots,m}_{n_m \text{ times}}, \ldots\rangle. \tag{S8}$$

For a symmetrized state it is not important which particles are in the $m^{\text{th}}$ level, but only how many. The factor $\sqrt{\prod_m n_m!}$ ensures the correct normalization, $\langle \boldsymbol{n}'|\boldsymbol{n}\rangle = \delta_{\boldsymbol{n},\boldsymbol{n}'}$, as can be checked with a straightaway calculation.

Consider now an operator $\hat{O} = \sum_{i=1}^N \hat{o}_i$ sum of identical single particle operators $\hat{o}_i$. In the chosen single particle basis, $\hat{o}_i$ reads

$$\hat{o}_i = \sum_{m,m'} O_{m',m} |m'\rangle_i \langle m|_i, \tag{S9}$$

with single-body matrix element $O_{m',m} = \langle m'|_i \hat{o}_i |m\rangle_i$. We are interested in how the many-body operator $\hat{O}$ acts on the symmetrized states $|\boldsymbol{n}\rangle$. Since $[\hat{S}, \hat{O}] = 0$, we have

$$\hat{O}|\boldsymbol{n}\rangle = \frac{1}{\sqrt{\prod_m n_m!}} \hat{S}\hat{O} |\underbrace{1,1,\ldots,1}_{n_1 \text{ times}}, \underbrace{2,2,\ldots,2}_{n_2 \text{ times}}, \ldots, \underbrace{m,m,\ldots,m}_{n_m \text{ times}}, \ldots\rangle, \tag{S10}$$

and so

$$\hat{O}|\boldsymbol{n}\rangle = \frac{1}{\sqrt{\prod_m n_m!}} \sum_m \sum_{m' \neq m} O_{m',m} \sum_{i=1}^N \hat{S}|m'\rangle_i \langle m|_i \left|\underbrace{1,1,\ldots,1}_{n_1 \text{ times}}, \underbrace{2,2,\ldots,2}_{n_2 \text{ times}}, \ldots, \underbrace{m,m,\ldots,m}_{n_m \text{ times}}, \ldots\right\rangle$$
$$+ \frac{1}{\sqrt{\prod_m n_m!}} \sum_m O_{m,m} \sum_{i=1}^N \hat{S}|m\rangle_i \langle m|_i |\underbrace{1,1,\ldots,1}_{n_1 \text{ times}}, \underbrace{2,2,\ldots,2}_{n_2 \text{ times}}, \ldots, \underbrace{m,m,\ldots,m}_{n_m \text{ times}}, \ldots\rangle, \tag{S11}$$

where we split diagonal and off diagonal terms. In the rightmost sum of both, there are $n_m$ non-vanishing terms, that is, those corresponding to the particles in state $m$. After the action of $\hat{S}$, these terms are all equal, and so

$$\hat{O}|\boldsymbol{n}\rangle = \frac{1}{\sqrt{\prod_m n_m!}} \sum_m \sum_{m' \neq m} O_{m',m} n_m \hat{S} |\ldots, \underbrace{m,m,\ldots,m}_{n_m-1 \text{ times}}, \ldots, \underbrace{m',m',\ldots,m'}_{n_{m'}+1 \text{ times}}, \ldots\rangle + \sum_m O_{m,m} n_m |\boldsymbol{n}\rangle. \tag{S12}$$

In the last expression, the first term also gives symmetrized states, but with different occupation numbers, and therefore involving a correction in the normalization

$$\hat{O}|\boldsymbol{n}\rangle = \sum_m \sum_{m' \neq m} O_{m',m} \sqrt{n_m(n_{m'}+1)} |n_1, \ldots, n_m - 1, \ldots, n_{m'} + 1, \ldots\rangle + \sum_m O_{m,m} n_m |\boldsymbol{n}\rangle, \tag{S13}$$

where we recognize the factors distinctive of bosonic operators. Indeed, we can formally define a pair of creation and anihilation bosonic operators $(b_m^\dagger, b_m)$ per each single-particle state $|m\rangle$, and write

$$\hat{O}|\boldsymbol{n}\rangle = \sum_{m,m'} O_{m',m} \hat{b}_{m'}^\dagger \hat{b}_m |\boldsymbol{n}\rangle. \tag{S14}$$

That is, calling $\hat{\boldsymbol{b}} = (\hat{b}_1; \hat{b}_2; ...)$ the column vector of bosonic annihilation operators, and $\hat{\boldsymbol{b}}^\dagger = (\hat{b}_1^\dagger, \hat{b}_2^\dagger, ...)$ the row vector of the bosonic creation operators, we can write

$$\hat{O} = \sum_i \hat{o}_i = \hat{\boldsymbol{b}}^\dagger \boldsymbol{O} \hat{\boldsymbol{b}}, \tag{S15}$$

where $\boldsymbol{O}$ is the matrix with single-particle entries $O_{m,n} = \langle m|_i \hat{o}_i |n\rangle_i$ (that is, a matrix of numbers). Symmetric states can be expressed in terms of bosonic operators as

$$|\boldsymbol{n}\rangle = \frac{\left(\hat{b}_1^\dagger\right)^{n_1}}{n_1!} \frac{\left(\hat{b}_2^\dagger\right)^{n_1}}{n_1!} ... \frac{\left(\hat{b}_m^\dagger\right)^{n_m}}{n_m!} ... |0\rangle. \tag{S16}$$

This derivation shows that, as long as the system is in the symmetrized sector, it is exactly equivalent to one of $N$ bosons in bosonic modes $\hat{b}_{m'}^\dagger$, associated with the single-particle states. This derivation is essentially that of second bosonization, but generalized to states of particles that are not necessarily bosons, but for which the wavefunction, Hamiltonian, and observables of interest can be considered symmetric under any permutation.

## Section S3: Preparation stage

In this Section, we consider the $t < 0$ preparation stage in which correlations in the many-body atomic system are induced through interactions or superradiance. The resulting atomic state at time $t = 0$ acts as the initial condition for the subsequent HHG process. We focus on the case in which such initialization protocol only involves the single-particle ground and first excited states, $|g\rangle$ and $|e\rangle$, respectively. To describe these, we introduce standard Pauli operators $\hat{\sigma}_{x,i}$, $\hat{\sigma}_{y,i}$, and $\hat{\sigma}_{z,i}$ for the $i$-th atom

$$\hat{\sigma}_{x,i} = |e\rangle_i \langle g|_i + |g\rangle_i \langle e|_i; \quad \hat{\sigma}_y = \sqrt{-1}(|g\rangle_i \langle e|_i - |e\rangle_i \langle g|_i); \quad \hat{\sigma}_z = |e\rangle_i \langle e|_i - |g\rangle_i \langle g|_i. \tag{S17}$$

Collective spin operators $\hat{S}_\nu$, with $\nu = x, y, z$, read

$$\hat{S}_\nu = \sum_{i=1}^N \hat{\sigma}_{\nu,i} = (\hat{b}_1^\dagger, \hat{b}_2^\dagger) \sigma_\nu \begin{pmatrix} \hat{b}_1 \\ \hat{b}_2 \end{pmatrix}, \tag{S18}$$

where the (non-hatted) $\sigma_\nu$ denotes the standard Pauli matrices. Raising and lowering operators are also introduced as $\hat{S}^\pm = \hat{S}_x \pm i\hat{S}_y$. With this notation at hand, we can now consider two separate preparation stages, one exploiting interatomic interactions and one exploiting superradiance.

**Preparation through $\pi/2$ pulse and shear force**

The first protocol we consider closely follows the one introduced in the context of spin squeezing by Kitagawa and Ueda [7], exploiting one-axis twisting. The collective and undriven Hamiltonian within the two lowest levels reads

$$\widehat{H} = \frac{\omega_0}{2} \hat{S}_z + \frac{\omega_J}{N} \hat{S}_z^2, \tag{S19}$$

where $\omega_0$ is the single body energy gap between $|g\rangle$ and $|e\rangle$, whereas $\omega_J$ is an energy scale associated to the atom-atom interaction (the factor $N$ guarantees extensivity of the Hamiltonian). First, at time $t = -t_h$, we rotate the spins with a $\pi/2$ pulse, leading them in a product state $|\Rightarrow\rangle$ in which all the spins are aligned along the direction $x$, $\hat{S}_x|\Rightarrow\rangle = N|\Rightarrow\rangle$. Then, we let the undriven Hamiltonian act on the system for a hold time $t_h$. The resulting wavefunction at time $t = 0$, that is the initial condition for HHG, reads

$$|\psi(0)\rangle = e^{-i\left(\frac{\omega_0}{2}\hat{S}_z + \frac{\omega_J}{N}\hat{S}_z^2\right)}|\Rightarrow\rangle. \tag{S20}$$

Conveniently, the time evolution operator $e^{-i\left(\frac{\omega_0}{2}\hat{S}_z + \frac{\omega_J}{N}\hat{S}_z^2\right)t_h}$ is diagonal in the chosen computational basis, which makes $|\psi(0)\rangle$ particularly easy to evaluate numerically. In Figure S1, the state $|\psi(0)\rangle$ is visualized for various hold times $t_h$ in terms of the atomic Wigner function on the Bloch sphere.

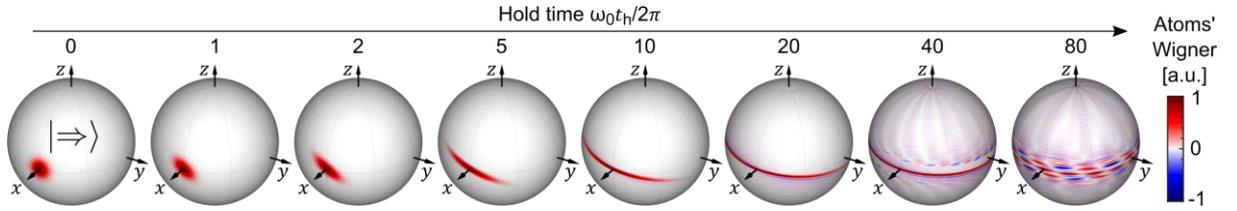

**Figure S1. Preparation through one-axis twisting**. The atoms can be prepared in a correlated state through interatomic $\hat{S}_z^2$ interaction. The state of the system versus time during this preparation stage is here visualized in terms of atomic Wigner function on the Bloch sphere. At time $t_h = 0$, the system is in $x$-polarized product state $|\Rightarrow\rangle$. The force $\hat{S}_z^2$ acts "shearing" the Wigner function, that progressively wraps around the Bloch sphere embracing its equator. At long enough times $t_h$, the Wigner function develops more pronounced negative parts. Here, we considered $N = 100$, $\omega_0 = 0.495$, and $\omega_J = 0.01$.

**Preparation through superradiance**

The second preparation protocol we consider is instead based on superradiance. First, at time $t = -t_h$ we act with a $\pi$-pulse to bring the system from the ground state $|\Downarrow\rangle$ to the excited state $|\Uparrow\rangle$. From the latter, the system then undergoes collective spontaneous emission, i.e., superradiance, that we describe with a standard Lindblad master equation [8,9] for the system's density matrix $\hat{\rho}$

$$\frac{d\hat{\rho}}{dt} = -i[\hat{H}, \hat{\rho}] + \gamma\left(\hat{S}^-\hat{\rho}\hat{S}^+ - \frac{1}{2}\{\hat{S}^+\hat{S}^-, \hat{\rho}\}\right), \tag{S21}$$

with $\gamma$ the decay rate. In the considered computational basis, for the considered initial condition $\hat{\rho}(-t_h) = |\Uparrow\rangle\langle\Uparrow|$, the density matrix $\hat{\rho}$ throughout the superradiance process is diagonal. Indeed, $\hat{\rho}$ commutes with anything diagonal, and in particular with functions of $\hat{S}_z$ like $\hat{S}^+\hat{S}^- = N(N+2) - \hat{S}_z^2 + 2\hat{S}_z$. We can therefore simplify Eq. (S21) to

$$\frac{d\hat{\rho}}{dt} = \gamma\left(\hat{S}^-\hat{\rho}\hat{S}^+ - \left(N(N+2) - \hat{S}_z^2 + 2\hat{S}_z\right)\hat{\rho}\right), \tag{S22}$$

which can conveniently be solved along its diagonal only. Eq. (S22) gives the well-known magnetization and emission intensity profiles for superradiance [10]. Alternatively, in Figure S2 we illustrate the superradiance from a less common perspective, that is, by plotting the atomic Wigner function on the Bloch sphere at various times.

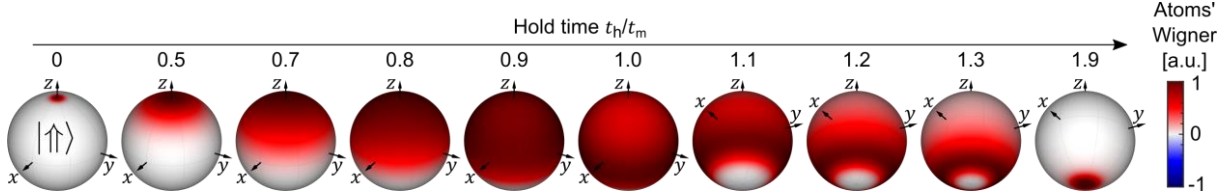

**Figure S2. Preparation through superradiance**. The atoms can be prepared in a correlated state through superradiance. The $t = 0$ state of the system versus time during this preparation stage is here visualized in terms of atomic Wigner function on the Bloch sphere. For vanishing hold times, that is $t_h = 0$, the system is in the product state $|\Uparrow\rangle$. For a hold time $t_h = t_m$, with $t_m$ the time of maximum superradiant emission, the Wigner appears rather uniform on the Bloch sphere (in agreement with [10]). For a hold time $t_h \approx 1.2 t_m$, the Wigner is concentrated on a ring lying between the equator and the south pole. For even longer hold times, the state of the system tends to $|\Downarrow\rangle$, and the Wigner concentrates on the South pole. Here, we considered $N = 100$ and $\gamma N = 0.1$.

## Section S4: Bosonic representation of strongly driven atoms

In this Section, we apply the bosonic representation of symmetric states introduced in Section 2 to describe strongly driven atoms. Indeed, under the assumptions of our model, the system under consideration is fully symmetric, and therefore amenable to this technique. The Hamiltonian reads (in atomic units):

$$\hat{H} = \sum_i^N \hat{h}_i - \left(E_c(t) + \hat{E}_q\right) \sum_i \hat{x}_i + \sum_{\boldsymbol{k}\sigma} \omega_k \hat{a}^\dagger_{\boldsymbol{k}\sigma} \hat{a}_{\boldsymbol{k}\sigma}, \tag{S23}$$

where $\hat{h}_i$ is the one-dimensional Hamiltonian of $i^{\text{th}}$ atom in Eq. (S1), $E_c(t)$ is the classical time-dependent field (linearly polarized along $x$), and $\hat{a}_{\boldsymbol{k}\sigma}$ and $\hat{a}^\dagger_{\boldsymbol{k}\sigma}$ the annihilation and creation operators, respectively, of photons with wavevector $\boldsymbol{k}$ and polarization $\sigma$. The quantized field reads

$$\hat{E}_q = \sum_{\boldsymbol{k}\sigma} \varepsilon_{\boldsymbol{k}\sigma}\left(g_{\boldsymbol{k}\sigma}\hat{a}_{\boldsymbol{k}\sigma} + g^*_{\boldsymbol{k}\sigma}\hat{a}^\dagger_{\boldsymbol{k}\sigma}\right), \qquad g_{\boldsymbol{k}\sigma} = i\sqrt{\frac{2\pi\omega_k}{V}}, \tag{S24}$$

with $\varepsilon_{\boldsymbol{k}\sigma}$ the projection in the $x$ direction of the polarization vector of the mode $(\boldsymbol{k}, \sigma)$ and $V \to \infty$ the volume. Using the bosonic representation in Section 2, taken with respect to the single-atom energy eigenstates $|m\rangle$ in Eq. (S4), we rewrite the Hamiltonian in Eq. (S23) as

$$\hat{H} = \hat{\boldsymbol{b}}^\dagger \boldsymbol{W} \hat{\boldsymbol{b}} - \left(E_c(t) + \hat{E}_q\right)\hat{\boldsymbol{b}}^\dagger \boldsymbol{D} \hat{\boldsymbol{b}} + \sum_{\boldsymbol{k}\sigma} \omega_k \hat{a}^\dagger_{\boldsymbol{k}\sigma} \hat{a}_{\boldsymbol{k}\sigma}, \tag{S25}$$

where $\boldsymbol{W}$ is a diagonal matrix of single-atom eigenenergies with entries $W_{nm} = \delta_{nm}w_n$, $\boldsymbol{D}$ is a dipole matrix with entries $D_{mn} = \langle m|\hat{x}|n\rangle$, and $\hat{\boldsymbol{b}} = (\hat{b}_1, \hat{b}_2, \dots, \hat{b}_m, \dots)$ is the vector of bosonic operators creating particles in the single-atom energy eigenstates.

The Heisenberg equations associated to the Hamiltonian (S25) read

$$\begin{cases} \dfrac{d\widehat{\boldsymbol{b}}}{dt} = -i\boldsymbol{W}\widehat{\boldsymbol{b}} + i\big(E_c(t) + \widehat{E}_q\big)\boldsymbol{D}\widehat{\boldsymbol{b}}, \\ \dfrac{d\hat{a}_{\boldsymbol{k}\sigma}}{dt} = -i\omega_k \hat{a}_{\boldsymbol{k}\sigma} + ig^*_{\boldsymbol{k}\sigma}\varepsilon_{\boldsymbol{k}\sigma}\widehat{\boldsymbol{b}}^\dagger \boldsymbol{D}\widehat{\boldsymbol{b}}. \end{cases} \quad (S26)$$

The second equation in Eq. (S26) is readily solved by

$$\hat{a}_{\boldsymbol{k}\sigma}(t) = \hat{a}_{\boldsymbol{k}\sigma}(0) + ig^*_{\boldsymbol{k}\sigma}\varepsilon_{\boldsymbol{k}\sigma}\int_{-\infty}^{t} d\tau\, e^{i\omega_k(\tau-t)}\widehat{\boldsymbol{b}}^\dagger(\tau)\boldsymbol{D}\widehat{\boldsymbol{b}}(\tau). \quad (S27)$$

To solve the first equation, instead, we note that the quantum part of the field $\widehat{E}_q$ can in first approximation be neglected with respect to the classical strong drive $E_c(t)$, yielding

$$\frac{d\widehat{\boldsymbol{b}}}{dt} = -i(\boldsymbol{W} - E_c(t)\boldsymbol{D})\widehat{\boldsymbol{b}}. \quad (S28)$$

Note that neglecting the photon back-action term $\widehat{E}_q$ is equivalent to neglecting superradiance. Indeed, we account for superradiance within the preparation stage (Section S3), and then we can neglect it during the strong drive pulse $E_c(t)$. Similar considerations hold for the atom-atom interactions, which are substantial during the slower preparation stage and negligible during the strong drive pulse. Eq. (S28) is linear in $\widehat{\boldsymbol{b}}$, and thus solved by

$$\widehat{\boldsymbol{b}}(t) = \boldsymbol{F}(t)\widehat{\boldsymbol{b}}(0), \quad (S30)$$

where $\widehat{\boldsymbol{b}}(0)$ is the vector of bosonic operators at time $t = 0$, when the strong pulse begins. The transition matrix is formally written as $\boldsymbol{F}(t) = T\exp\left(-i\int_0^t (\boldsymbol{W} - E_c(\tau)\boldsymbol{D})d\tau\right)$, with $T$ the time-ordering operator. We emphasize that, while $\widehat{\boldsymbol{b}}$ is a vector of abstract operators, each of them living in an exponentially large Hilbert space, $\boldsymbol{W}$, $\boldsymbol{D}$, and $\boldsymbol{F}(t)$ are comparably much simpler objects, namely $(M \times M)$-dimensional matrices of numbers (not operators!), that can be easily stored on a computer. These practical computational issues will be emphasized in the summary and hands-on Section 6.

Knowing the solution in Eq. (S22), the time evolution of any bilinear operator $\widehat{O} = \widehat{\boldsymbol{b}}^\dagger \boldsymbol{O}\widehat{\boldsymbol{b}}$ is readily found as

$$\widehat{O}(t) = \widehat{\boldsymbol{b}}^\dagger(t)\boldsymbol{O}\widehat{\boldsymbol{b}}(t) = \widehat{\boldsymbol{b}}^\dagger(0)\boldsymbol{F}^\dagger(t)\boldsymbol{O}\boldsymbol{F}(t)\widehat{\boldsymbol{b}}(0).$$

Crucially, since all the preparation procedures of the initial state we consider only involve the lowest two levels, meaning $\hat{b}_m^\dagger \hat{b}_m |\psi(0)\rangle = 0$ for any $m > 2$, we can in practice substitute $\widehat{\boldsymbol{b}}^\dagger(0) \to \left(\hat{b}_1^\dagger(0), \hat{b}_2^\dagger(0)\right)$ and $\widehat{\boldsymbol{b}}(0) \to \begin{pmatrix}\hat{b}_1(0) \\ \hat{b}_2(0)\end{pmatrix}$, which will be from now on implicit in our notation, and reduce the expression of $\widehat{O}(t)$ to

$$\widehat{O}(t) = \widehat{\boldsymbol{b}}^\dagger(0)\boldsymbol{o}(t)\widehat{\boldsymbol{b}}(0), \quad (S31)$$

where $\boldsymbol{o}(t)$ is the two-by-two top left corner of $\boldsymbol{F}^\dagger(t)\boldsymbol{O}\boldsymbol{F}(t)$, that is

$$\boldsymbol{o}(t) = \begin{pmatrix} \left(\boldsymbol{F}^\dagger(t)\boldsymbol{O}\boldsymbol{F}(t)\right)_{11} & \left(\boldsymbol{F}^\dagger(t)\boldsymbol{O}\boldsymbol{F}(t)\right)_{12} \\ \left(\boldsymbol{F}^\dagger(t)\boldsymbol{O}\boldsymbol{F}(t)\right)_{21} & \left(\boldsymbol{F}^\dagger(t)\boldsymbol{O}\boldsymbol{F}(t)\right)_{22} \end{pmatrix}. \quad (S32)$$

In practice, therefore, we can numerically evolve $\boldsymbol{F}(t)$ while only saving the much smaller two-by-two matrix $\boldsymbol{o}(t)$ in the computer memory, for each observable of interest. While it might be naively perceived as a somewhat "two-level matrix", $\boldsymbol{o}(t)$ contains the full many-level information on the dynamics, because obtained as a subpart of the many-level matrix $\boldsymbol{F}^\dagger(t)\boldsymbol{O}\boldsymbol{F}(t)$. The matrix $\boldsymbol{o}(t)$, which we might call the *dynamical matrix* of the operator $\hat{O}$, is a key object in our theory. Further, we can write more explicitly

$$\hat{O}(t) = o_{11}(t)\hat{b}_1^\dagger(0)\hat{b}_1(0) + o_{12}(t)\hat{b}_1^\dagger(0)\hat{b}_2(0) + o_{21}(t)\hat{b}_2^\dagger(0)\hat{b}_1(0) + o_{22}(t)\hat{b}_2^\dagger(0)\hat{b}_2(0), \quad (S33)$$

which, exploiting $\hat{S}_\nu(0) = \hat{\boldsymbol{b}}^\dagger(0)\sigma_\nu\hat{\boldsymbol{b}}(0)$ and $N = \hat{\boldsymbol{b}}^\dagger(0)\hat{\boldsymbol{b}}(0)$, can be conveniently rewritten in terms of collective spin operators

$$\hat{O}(t) = \frac{o_{11}(t) + o_{22}(t)}{2}N + \frac{o_{22}(t,t_1) - o_{11}(t,t_1)}{2}\hat{S}_z(0) + o_{21}(t)\hat{S}^+(0) + o_{12}(t)\hat{S}^-(0). \quad (S34)$$

That is, any operator $\hat{O}(t) = \sum_i \hat{o}_i$ at time $t$ can be written as a linear combination of standard collective spin (i.e., angular momentum) matrices, provided that the respective dynamical matrix $\boldsymbol{o}(t)$ has been computed.

Central among the observables of interest is the position, or dipole moment, $\sum_i \hat{x}_i = \hat{\boldsymbol{b}}^\dagger \boldsymbol{D}\hat{\boldsymbol{b}}$. We compute its two-by-two dynamical matrix $\boldsymbol{d}(t)$ as in Eq. (S32) and use Eqs. (S27) and (S31) to compactly write the time-evolved light field operators at long times (after the pulse)

$$\hat{a}_{\boldsymbol{k}\sigma} = \hat{a}_{\boldsymbol{k}\sigma}(0) + ig_{\boldsymbol{k}\sigma}^* \varepsilon_{\boldsymbol{k}\sigma}\hat{\boldsymbol{b}}^\dagger(0)\tilde{\boldsymbol{d}}(\omega_k)\hat{\boldsymbol{b}}(0), \quad (S35)$$

where $\tilde{\boldsymbol{d}}(\omega) = \int_0^{+\infty} dt\, e^{i\omega t}\boldsymbol{d}(t)$ is the one-sided Fourier transform of the dynamical matrix of the dipole moment (thus, a $\omega$-dependent $2\times 2$ matrix).

The expression in Eq. (S35) can be for instance used to compute the average number of emitted photons in the frequency window $(\omega - d\omega/2, \omega + d\omega/2)$, reading

$$N_\omega = \sum_{\boldsymbol{k}\sigma:|\omega_k-\omega|<\frac{d\omega}{2}} \langle \hat{a}_{\boldsymbol{k}\sigma}^\dagger \hat{a}_{\boldsymbol{k}\sigma}\rangle \approx \sum_{\boldsymbol{k}\sigma:|\omega_k-\omega|<\frac{d\omega}{2}} \left\langle \varepsilon_{\boldsymbol{k}\sigma}^2 |g_{\boldsymbol{k}\sigma}|^2 \left(\hat{\boldsymbol{b}}^\dagger(0)\tilde{\boldsymbol{d}}^\dagger(\omega_k)\hat{\boldsymbol{b}}(0)\right)\left(\hat{\boldsymbol{b}}^\dagger(0)\tilde{\boldsymbol{d}}(\omega_k)\hat{\boldsymbol{b}}(0)\right)\right\rangle$$

$$\approx \frac{2}{3}\frac{\omega^3}{\pi c^3}\left\langle \sum_{i,j,k,l=1,2} \hat{b}_i^\dagger(0)\hat{b}_j(0)\hat{b}_k^\dagger(0)\hat{b}_l(0)\, \tilde{d}_{ij}^{\,\dagger}(\omega)\tilde{d}_{kl}(\omega)\right\rangle d\omega. \quad (S36)$$

In the last expression, we have neglected terms involving hopping to higher levels $m > 2$. Indeed, because the initial condition only involves the two lowest levels, the terms coming from $m > 2$ are negligible (being them smaller by at least a factor $1/N$ as compared to the ones that we retained). The spectrum is readily found from Eq. (S36)

$$\frac{d\epsilon}{d\omega} = \frac{\omega N_\omega}{d\omega} = \frac{2}{3}\frac{\omega^4}{\pi c^3}\left\langle \sum_{i,j,k,l=1,2} \hat{b}_i^\dagger(0)\hat{b}_j(0)\hat{b}_k^\dagger(0)\hat{b}_l(0)\, \tilde{d}_{ij}^{\,\dagger}(\omega)\tilde{d}_{kl}(\omega)\right\rangle. \quad (S37)$$

For instance, in the simplest scenario in which the atoms are initially all in the ground state, the dominant term in the sum is that with $i = j = k = l = 1$, yielding the familiar expression [1]

$$\frac{d\epsilon}{d\omega} = N^2 \frac{2}{3} \frac{\omega^4}{\pi c^3} |\tilde{d}_{11}(\omega)|^2. \tag{S38}$$

However, the modes $a_{k\sigma}$ in Eq. (S35) contain much more information than that on the emission spectrum. To debunk it, we want to define a mode associated with the emission into a solid angle window $d\Omega$ of a detector, with a given polarization (for which, within $d\Omega$, we can consider $\varepsilon_{k\sigma} \approx 1$ for one of the two polarizations, on which we focus) and within the frequency window $\left(n\omega_d - \frac{d\omega}{2}, n\omega_d + \frac{d\omega}{2}\right)$ around the $n$-th harmonics (Figure S3). The number of photons fulfilling these constraints can be found similarly as for Eq. (S36), and reads

$$N_n \approx d\omega \frac{d\Omega}{4\pi} \frac{(n\omega_d)^3}{\pi c^3} \left\langle \sum_{i,j,k,l=1,2} \hat{b}_i^\dagger(0)\hat{b}_j(0)\hat{b}_k^\dagger(0)\hat{b}_l(0)\, \tilde{d}_{ij}^{\,\dagger}(n\omega_d)\tilde{d}_{kl}(n\omega_d) \right\rangle. \tag{S39}$$

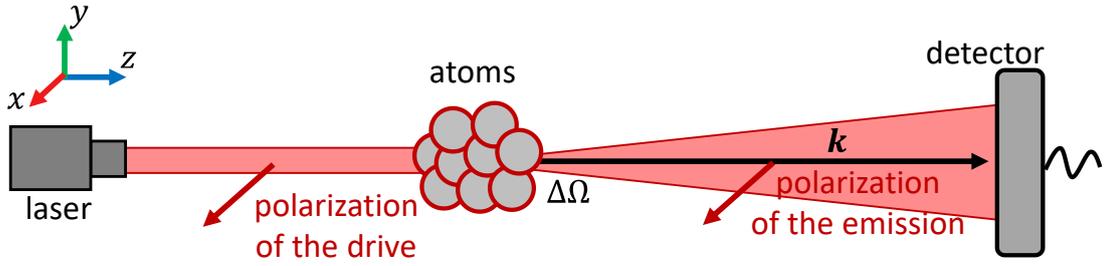

**Figure S3. Schematic of the setup for HHG**. We consider the polarization of the emitted light parallel to the initial driving field polarization. The emitted light is considered only within a small solid angle $\Delta\Omega$.

We can thus define a mode $\hat{a}_n$ associated to a given harmonic $n$ as

$$\hat{a}_n = \frac{1}{\sqrt{\mathcal{N}}} \sum_{\mathbf{k} \in n} \hat{a}_{\mathbf{k}\sigma}, \tag{S40}$$

where the notation $\mathbf{k} \in n$ means that the sum runs over the modes fulfilling the selection criteria above, whereas norm is a not yet specified normalization factor. Plugging Eq. (S35) into Eq. (S40), we get

$$\hat{a}_n = \hat{a}_n(0) + \hat{\mathbf{b}}^\dagger(0) \mathbf{d}_n \hat{\mathbf{b}}(0), \tag{S41}$$

where we defined $\hat{a}_n(0) = \frac{1}{\sqrt{\mathcal{N}}} \sum_{\mathbf{k} \in n} \hat{a}_{\mathbf{k}\sigma}(0)$ and

$$\mathbf{d}_n = \frac{i}{\sqrt{\mathcal{N}}} \sum_{\mathbf{k} \in n} g_{\mathbf{k}\sigma}^* \tilde{\mathbf{d}}(\omega_k). \tag{S42}$$

We now choose the normalization factor $\mathcal{N}$ such that $[\hat{a}_n, \hat{a}_n^\dagger] \approx 1$. Not only does this choice render $\hat{a}_n$ a suitable photonic operator making it fulfill the standard commutation relations, but, as we will see, it also allows the correct estimate of the number of photons, which shows the consistency of this derivation. We have

$$[\hat{a}_n, \hat{a}_n^\dagger] = \frac{1}{\mathcal{N}} \sum_{k \in n} \sum_{k' \in n} [\hat{a}_{k\sigma}(0), \hat{a}_{k'\sigma}^\dagger(0)] + \hat{\boldsymbol{b}}^\dagger(0)[\boldsymbol{d}_n, \boldsymbol{d}_n^\dagger]\hat{\boldsymbol{b}}(0). \tag{S43}$$

Let us now assume a priori that the 2x2 dimensional matrix $\boldsymbol{d}_n$ has very small entries, such that $\hat{\boldsymbol{b}}^\dagger(0)[\boldsymbol{d}_n, \boldsymbol{d}_n^\dagger]\hat{\boldsymbol{b}}(0)$ can be neglected in Eq. (S36). Under this assumption, we have

$$[\hat{a}_n, \hat{a}_n^\dagger] = \frac{1}{\mathcal{N}} \sum_{k \in n} 1 \approx \frac{1}{\mathcal{N}} \frac{n^2 \omega_d^2 V d\Omega}{(2\pi)^3 c^3} d\omega, \tag{S44}$$

and, thus,

$$\mathcal{N} = \frac{n^2 \omega_d^2 V d\Omega}{(2\pi)^3 c^3} d\omega, \tag{S45}$$

and

$$\boldsymbol{d}_n \approx \sqrt{\frac{d\Omega}{4\pi} \frac{n^3 \omega_d^3}{\pi c^3} d\omega} \, \tilde{\boldsymbol{d}}(n\omega_d). \tag{S46}$$

The a priori assumption made to get from to Eq. (S43) to Eq. (S46) can now be checked as consistent. Indeed, in the relevant cases that we will consider we get from Eq. (S46) $|\boldsymbol{d}_n| \sim 10^{-4}$, so that, as long as $N \ll 10^8$, we have $\langle \hat{\boldsymbol{b}}^\dagger(0)[\boldsymbol{d}_n \boldsymbol{d}_n^\dagger]\hat{\boldsymbol{b}}(0) \rangle \ll 1$, as we assumed. As a further consistency check, we note that, computing the number of photons in the mode $N_n$ as $\hat{a}_n^\dagger \hat{a}_n$, the same result as in Eq. (S39) is found, which comes "for free" from having enforced $[\hat{a}_n, \hat{a}_n^\dagger] \approx 1$. In our calculations, we consider $d\omega = 0.5\omega_d$ and $d\Omega = \frac{4\pi}{3}$.

Having built a suitable mode $\hat{a}_n$ for the $n$-th harmonc, we can now use it to extract information on the emission, accessing to its quantum features beyond the bare average number of photons. Indeed, since $\hat{a}_n(0)|0\rangle = 0$, we can compute the normally ordered moments of $\hat{a}_n$ as

$$\left\langle \left(\hat{a}_n^\dagger\right)^m (\hat{a}_n)^l \right\rangle = \left\langle \left(\left(\hat{\boldsymbol{b}}^\dagger(0)\boldsymbol{d}_n\hat{\boldsymbol{b}}(0)\right)^\dagger\right)^m \left(\hat{\boldsymbol{b}}^\dagger(0)\boldsymbol{d}_n\hat{\boldsymbol{b}}(0)\right)^l \right\rangle. \tag{S47}$$

Therefore, for the sake of computing the expectation value of normally ordered powers of $\hat{a}_n$ and $\hat{a}_n^\dagger$, the photonic operator can be written omitting the vacuum term $\hat{a}_n(0)$, thus as

$$\hat{a}_n = \frac{(d_n)_{11} + (d_n)_{22}}{2} N + \frac{(d_n)_{22} - (d_n)_{11}}{2} \hat{S}_z(0) + (d_n)_{21}\hat{S}^+(0) + (d_n)_{12}\hat{S}^-(0). \tag{S48}$$

That is, once computed the $2 \times 2$ matrix $d_n$ (which is indeed the main computational challenge in our theory), the photonic operator $\hat{a}_n$ is just a simple linear combination of standard angular momentum operators, that can be easily represented as sparse $(N+1) \times (N+1)$ dimensional matrices on the computer (more details on the practical implementation of our theory will follow in Section 6).

# Section S5: From normally ordered moments to the Wigner function and photon statistics

In this Section, we show how to exploit the information contained in the expectation value of normally ordered moments $\langle (\hat{a}^\dagger)^m (\hat{a})^l \rangle$ to compute the Wigner function and photons statistics associated to the mode $\hat{a}$ (for us, the modes of interest will then be the harmonic ones, $\hat{a}_n$).

**Wigner function from moments**

We begin by writing the Wigner function in terms of the Glauber function [11]

$$W(\alpha) = \frac{1}{\pi} \int P(\beta) \exp(-2|\beta - \alpha|^2) \, d^2\beta. \tag{S49}$$

The Glauber representation is convenient because it allows a simple expression of normally ordered moments, namely [11]

$$\langle \hat{a}^{\dagger m} \hat{a}^l \rangle = \int P(\beta) (\beta^*)^m \beta^l d^2\beta. \tag{S50}$$

The exponent in Eq. (S49) can then be expanded as

$$e^{-2|\beta-\alpha|^2} = e^{-2|\alpha|^2 - 2\beta\beta^* + 2\beta^*\alpha + 2\beta\alpha^*} = e^{-2|\alpha|^2} \sum_{l,m,k} \frac{(-1)^l 2^{l+m+k} \alpha^m \alpha^{*k} \beta^{l+k} \beta^{*l+m}}{l! \, m! \, k!}, \tag{S51}$$

that, after a couple of changes of indices, yields

$$e^{-2|\beta-\alpha|^2} = e^{-2|\alpha|^2} \sum_{l,m,k} \frac{(-1)^{l-k} 2^{m+k} \alpha^{m-l} |\alpha|^{2k} \beta^l \beta^{*m}}{(l-k)! \, (m+k-l)! \, k!}. \tag{S52}$$

Plugging Eqs. (S50) and (S52) into Eq. (S49), we get

$$W(\alpha) = \frac{e^{-2|\alpha|^2}}{\pi} \sum_{l,m} (-1)^l 2^m \alpha^{m-l} \langle \hat{a}^{\dagger m} \hat{a}^l \rangle \left( \sum_k \frac{(-1)^k (2|\alpha|^2)^k}{(l-k)! \, (m+k-l)! \, k!} \right), \tag{S53}$$

that, by calling $\xi_{lm}(\alpha)$ all what is left apart from $\langle \hat{a}_n^{\dagger m} \hat{a}_n^l \rangle$, can be compactly written as

$$W(\alpha) = \sum_{m,l} \langle \hat{a}_n^{\dagger m} \hat{a}_n^l \rangle \xi_{lm}(\alpha). \tag{S54}$$

The functions $\xi_{lm}(\alpha)$ can be expressed in terms of the Kummer confluent hypergeometric function of the second kind $U$ as

$$\xi_{lm}(\alpha) = \frac{2 e^{-2|\alpha|^2}}{\pi} \alpha^{m-l} 2^m \frac{U(-l, m-l+1, 2|\alpha|^2)}{l! \, m!}. \tag{S55}$$

and fulfill the property $\xi_{lm}(\alpha) = \xi_{ml}(\alpha^*)$. Computing $\xi_{lm}(\alpha)$ is in general challenging: on the one hand, the sum over $k$ in Eq. (S53) makes the calculation long, whereas on the other hand the alternating sign from the term $(-1)^k$ makes it subject to numerical cancellation. To circumvent some of these challenges, we seek to compute $\xi_{lm}(\alpha)$ by recursion. Starting from known recursion relations for the Kummer function $U$ [12], we get

$$\xi_{lm} = -\frac{1}{\alpha l}\left(2\alpha^* \xi_{l-2,m} + (m - l + 1 - 2|\alpha|^2)\xi_{l-1,m}\right). \tag{S56}$$

Given for $l = 0,1$ that

$$\begin{cases} \xi_{0m}(\alpha) = \xi_{m0}(\alpha^*) = \dfrac{2e^{-2|\alpha|^2}}{\pi}\dfrac{2^m \alpha^m}{m!}, \\ \xi_{1,m}(\alpha) = \xi_{m,1}(\alpha^*) = \dfrac{2e^{-2|\alpha|^2}}{\pi}\dfrac{2^m \alpha^{m-1}}{m!}(2|\alpha|^2 - m). \end{cases} \tag{S57}$$

We can therefore efficiently reconstruct by recursion all the values of $\xi_{lm}$ and compute the Wigner function in Eq (S54).

**Photon statistics from the Wigner function**

With the Wigner function available, it is straightforward to obtain the full photon statistics. Indeed, the average of any operator $G(\alpha)$ [13] equals to

$$\langle G(\alpha)\rangle = \int d^2\alpha\, W(\alpha) g(\alpha), \tag{S58}$$

where $g(\alpha) = \int d\text{Im}[\alpha]\langle \sqrt{2}\text{Re}[\alpha] - \text{Im}[\alpha]/\sqrt{2}|G(\alpha)|\sqrt{2}\text{Re}[\alpha] + \text{Im}[\alpha]/\sqrt{2}\rangle e^{i\sqrt{2}\text{Im}[\alpha]y}$ is the Wigner symbol associated to the operator $G$. The probability of observing $n$ photons can be expressed as

$$p_n = \langle |n\rangle\langle n|\rangle, \tag{S59}$$

that is, as the expectation value of the projection operator $|n\rangle\langle n|$ onto the Fock state $|n\rangle$ with $n$ photons. The Wigner symbol of such an operator is easily found as

$$g_n(\alpha) = 4(-1)^n e^{-2|\alpha|^2} L_n(4|\alpha|^2), \tag{S60}$$

and so the final photon statistics reads

$$p_n = \int d^2\alpha\, W(\alpha) g_n(\alpha). \tag{S61}$$

**Joint photon statistics and correlations between different modes**

Here we propose an alternative derivation of $p_n$ as compared to the one above. Indeed, while such alternative is numerically less stable than Eq. (S61), it has the advantage that it can be readily generalized to compute the joint probability distribution $p_{nm}$, that is the probability of observing $n$ photons in a given harmonic, and $m$ photons in a second given harmonic.

To begin with, we express the moments $\langle \hat{a}^{\dagger m} \hat{a}^m \rangle$ to the single mode photon statistics $p_n$ (that is, the probability of observing $n$ photons in the mode $a$)

$$\langle \hat{a}^{\dagger m} \hat{a}^m \rangle = \sum_{n=m}^{\infty} p_n \frac{n!}{(n-m)!}. \tag{S62}$$

We want to find the coefficients $c_{lm}$ that reverse Eq. (S62) as

$$p_l = \sum_m c_{lm} \langle \hat{a}^{\dagger m} \hat{a}^m \rangle, \tag{S63}$$

This can be achieved substituting Eq. (S63) into Eq. (S62), from which we get the condition

$$\sum_{m=l}^n c_{lm} \frac{n!}{(n-m)!} = \delta_{nl}, \tag{S64}$$

with $\delta_{nl}$ the Kronecker delta. The system of linear equations (S64) is solved by

$$c_{lm} = \frac{(-1)^{m-l}}{l!\,(m-l)!}, \tag{S65}$$

that, plugged back into Eq. (S63), gives

$$p_n = \sum_m \frac{(-1)^{m-n}}{n!\,(m-n)!} \langle \hat{a}^{\dagger m} \hat{a}^m \rangle. \tag{S66}$$

Eq. (S61) and Eq. (S66) provide of course the same result, although we find that the former is easier to evaluate numerically, thanks to less severe involved numerical rounding errors for large number of photons.

Now let us calculate the joint probability $p_{nm}$ to have $n$ photons in a mode $\hat{a}_1$ and $m$ photons in a mode $\hat{a}_2$. Generalizing Eqs. (S62) and (S63) we have

$$\langle \hat{a}^{\dagger k}_1 \hat{a}^k_1 \hat{a}^{\dagger l}_2 \hat{a}^l_2 \rangle = \sum_{n,m} p_{nm} \frac{n!}{(n-k)!} \frac{m!}{(m-l)!}, \tag{S67}$$

and

$$p_{nm} = \sum_{i,j} c_{in} c_{jm} \langle \hat{a}^{\dagger i}_1 \hat{a}^i_1 \hat{a}^{\dagger j}_2 \hat{a}^j_2 \rangle. \tag{S68}$$

We substitute Eq. (S68) in Eq. (S67) and get a system of linear equations in the unknowns $c_{in}$

$$\sum_n c_{in} \frac{n!}{(n-k)!} = \delta_{ik}, \qquad \sum_m c_{jm} \frac{m!}{(m-l)!} = \delta_{jl}. \tag{S69}$$

Solving it as for Eq. (S64), we get to the final expression

$$p_{nm} = \sum_{k,l=m}^{\infty} \frac{(-1)^{k-n}}{n!\,(k-n)!} \frac{(-1)^{l-m}}{m!\,(l-m)!} \langle \hat{a}^{\dagger k}_1 \hat{a}^k_1 \hat{a}^{\dagger l}_2 \hat{a}^l_2 \rangle. \tag{S70}$$

# Section S6: Implementation oriented summary

The numerical implementation of the above theory unfolds along the following main steps, which can also serve as a summary.

1. **Single-particle spectrum**
   The first step is finding the spectrum of the single-particle Hamiltonian $\hat{h} = \frac{\hat{p}^2}{2} + V(\hat{x})$, Eq. (S4). The operators $\hat{x}$ and $\hat{p}$ are discretized and represented as $M \times M$ dimensional matrices, with $M \approx \frac{2L}{dx}$ (e.g., $M = 429$ for the parameters considered throughout the main text). Specifically, the position basis is chosen, so that $\hat{x}$ and $\hat{p}$ are represented as a diagonal and a tridiagonal matrix, respectively. The single-particle eigenstates $|m\rangle$, with $m = 1,2,\ldots,M$, are found via exact diagonalization of the discretized single atom Hamiltonian $\hat{h}$, and stored on the computer as $M$-components vectors. The dipole moment matrix $\boldsymbol{D}$ is build element by element, with $D_{nm} = \langle n|\hat{x}|m\rangle$, and is $M \times M$ dimensional. The diagonal matrix $\boldsymbol{W}$ of the single-particle eigenenergies is also built.

2. **Solution of the preparation stage (in Schrödinger picture)**
   The preparation stage within the two lowest atomic levels can be solved exactly. We consider the $(N+1)$-dimensional many-body Hilbert space of the symmetrized states of the two lowest single-particle eigenstates $|g\rangle$ and $|e\rangle$. Considering as a computational basis that in which $\hat{S}_z$ is diagonal, we build the $(N+1) \times (N+1)$-dimensional matrices of the collective spin operators $\hat{S}_{x,y,z}$. For the sake of more efficient computations, these operators (and the following ones) are represented as sparse matrices. We consider three different initial conditions: (i) trivial (uncorrelated) wavefunctions $|\psi(0)\rangle$, such as $|\Uparrow\rangle$, $|\Downarrow\rangle$, and $|\Rightarrow\rangle$, that can be directly represented as $(N+1)$-dimensional vectors, (ii) sheared states, obtained from Eq. (S20), (iii) states from superradiance, obtained in the form of a $(N+1) \times (N+1)$-dimensional diagonal density matrix $\hat{\rho}(0)$ by numerically integrating Eq. (S22). All these initial conditions can be represented in terms of an atomic Wigner function on a sphere [14].

3. **Integration of the HHG pulse**
   We numerically time evolve the evolution matrix $\boldsymbol{F}(t) = T\exp\left(-\frac{i}{\hbar}\int_{-\infty}^{t}(\boldsymbol{W} - E_c(\tau)\boldsymbol{D})d\tau\right)$ in Eq. (S30). This is a $M \times M$ dimensional matrix as $\boldsymbol{W}$ and $\boldsymbol{D}$ also are (therefore, unlike the vector of bosonic operators $\hat{\boldsymbol{b}}$, living in an abstract Hilbert space instead). This task breaks down to computing $\boldsymbol{F}(t + dt) = \exp[-i(\boldsymbol{W} - E_c(t)\boldsymbol{D})dt]\,\boldsymbol{F}(t)$, that is, the exponential of a matrix times another matrix, which can be done efficiently [15]. The full time history of $\boldsymbol{F}(t)$ is not stored in the memory: instead, we only save the $2 \times 2$ dynamical matrix $\boldsymbol{o}(t)$ for the observables of interest, and in particular, the matrix $\boldsymbol{d}(t)$ associated to the position operator, see Eqs. (S30) and (S32).

4. **Get the photonic operator for the harmonics of interest**
   Fourier transforming $\boldsymbol{d}(t)$ [Eq. (S35)] and multiplying it times the suitable coefficients [Eq. (S46)], we find the $2 \times 2$ matrix $\boldsymbol{d}_n$ for the harmonics $n$ of interest. From this, the photonic operator associated to the harmonic, $\hat{a}_n$, is obtained as a linear combination of collective spin operators [Eq. (S48)]. By doing this, we represent the photonic operator $\hat{a}_n$ on our computer in the form of a $(N+1) \times (N+1)$-dimensional matrix. Such an operator misses the vacuum term, see Eq. (S48), under the assumption that it is only used to compute normally ordered moments.

5. **Compute normally ordered moments**
   The moments $\langle \hat{a}_n^{\dagger\,m} \hat{a}_n^l \rangle$ are computed for the desired initial conditions, e.g. $\langle \hat{a}_n^{\dagger\,m} \hat{a}_n^l \rangle = \mathrm{Tr}\left(\hat{\rho}(0)\hat{a}_n^{\dagger\,m}\hat{a}_n^l\right)$ in the case of preparation through superradiance, or $\langle \hat{a}_n^{\dagger\,m} \hat{a}_n^l \rangle =$ $\langle\psi(0)|\hat{a}_n^{\dagger m}\hat{a}_n^l|\psi(0)\rangle$ in the case of a pure initial state $|\psi(0)\rangle$. For these operations, we keep working in the $(N+1)\times(N+1)$-dimensional Hilbert space.

6. **Compute the photonic Wigner function**
   From the normally ordered moments, we reconstruct the photonic Wigner function as in Eqs. (S54), (S56), and (S57). Note: to facilitate numerical convergence in this step, it is advised to perform a shift $\hat{a}_n \to \hat{a}_n - \langle\hat{a}_n\rangle$.
7. **Compute the photon statistics**
   From the Wigner function, we extract the photons statistics as in Eq. (S61).

## Section S7: Complementary results

In this section, we report a few complementary results expanding those of the main text.

In Figures S4 and S5, we consider the same scenarios as in Figures 3 and 4 in the main text, respectively, but plot the photon statistics rather than the Wigner function. While the photon statistics contains in general les information than the Wigner function, it is on the other hand a simpler diagnostics to measure and quantify the deviations of the emitted light from a coherent state.

In Figure S6, we report the photon number statistics as in Figure 2 in the main, but adding the column corresponding to the initial condition $|\psi(0)\rangle = |\Uparrow\rangle$, that can be obtained from the ground state $|\Downarrow\rangle$ with a coherent preparation $\pi$ pulse. This highlights how the number of emitted photons (that is, the efficiency of HHG) strongly depends not only on the considered harmonics, but on the atomic initial condition, too. Indeed, for $|\psi(0)\rangle = |\Uparrow\rangle$ we find, for the considered harmonics, a significantly lower number of emitted photons as compared to $|\psi(0)\rangle = |\Downarrow\rangle, |\Rightarrow\rangle$, and $|N/2\rangle$.

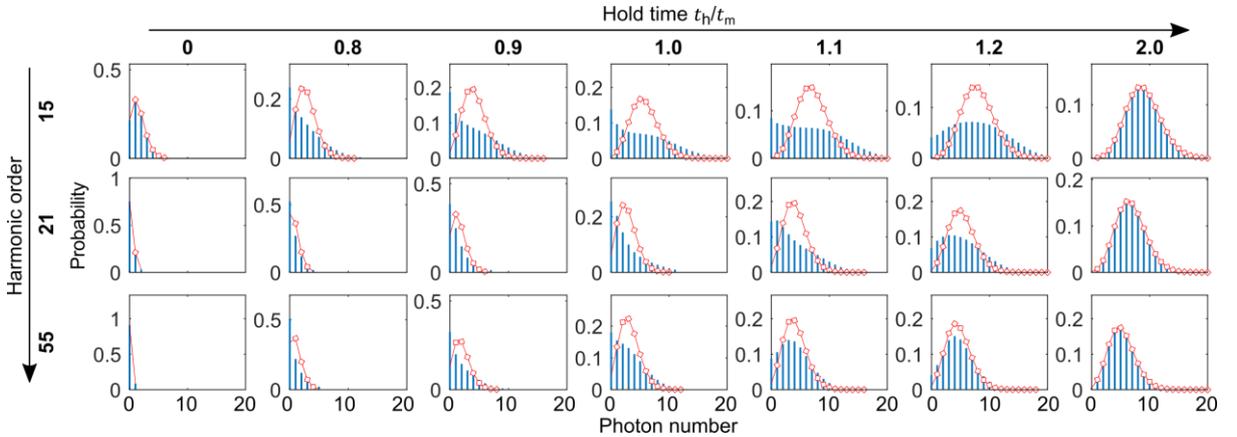

**Figure S4. Photon statistics from strongly driven superradiant atoms**. We consider the same scenario and parameters as in Figure 4c in the main text, but plot the photon statistics instead. To help appreciate the deviation from a coherent state, we plot for comparison a Poissonian fit in red. We observe that, for $t_h = 0$, that is for atoms initially all in their first excited state $|\psi(0)\rangle = |\Uparrow\rangle$, the emission is particularly inefficient, and fewer photons are emitted as compared, for instance, to the case of $t_h \to \infty$ (that is, ground state HHG with $|\psi(0)\rangle = |\Downarrow\rangle$).

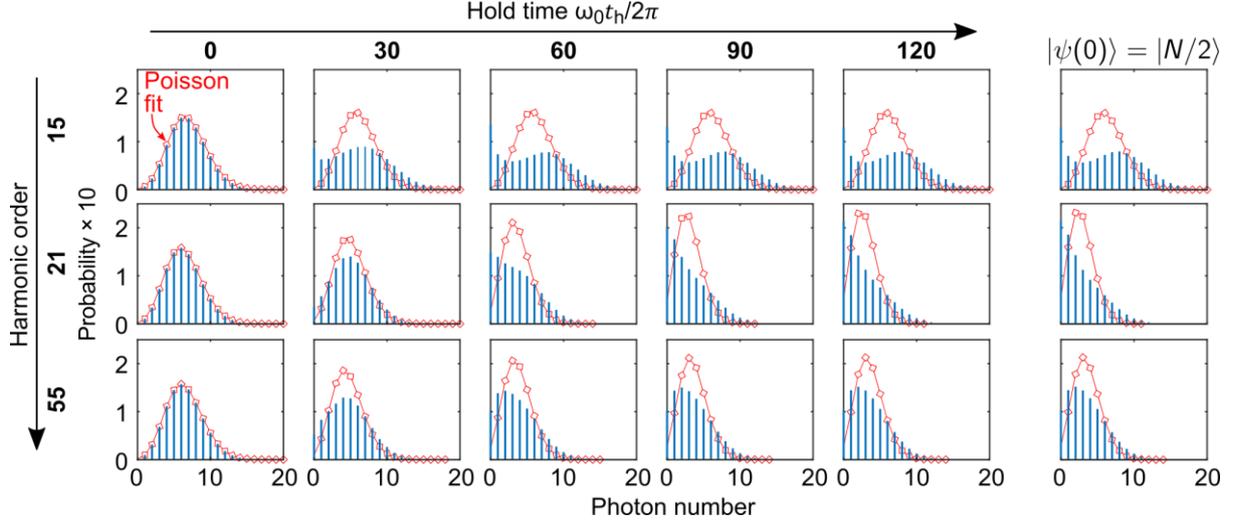

**Figure S5. Photon statistics from strongly driven interacting atoms.** We consider the same scenario and parameters as in Figure 3c in the main text, but plot the photon statistics instead. To help appreciate the deviation from a coherent state, we plot for comparison a Poissonian fit in red.

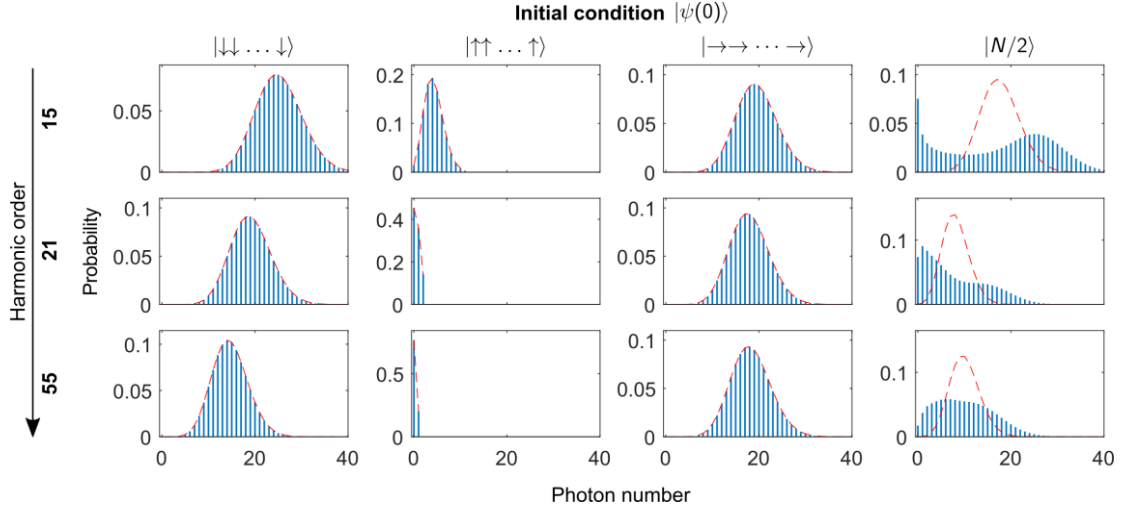

**Figure S6. Dependence of the HHG efficiency on the initial condition.** We plot the photon number statistics as in Figure 2 in the main text, but with an additional column (the second from left) presenting the case of an initial condition with all atoms in an excited state (marked as $|\Uparrow\rangle$). The emission from such an initial atomic state is weaker (lower photon numbers in the presented harmonics), which emphasizes the strong dependence of the HHG efficiency on the initial condition.

## Section S8: Phase-space approach

In Section S4 we obtained equations describing the atoms' dynamics [Eqs. (S26) and (S28)] and the emitted radiation [Eqs. (S26) and (S27)]. The solution of these equations was provided in Eqs. (S30) and (S41), and delegates most of the numerical challenges to the computation of the transition matrix $F$, that accounts for the single-particle atomic spectrum. These equations deal with operators, namely the bosonic operators in $\hat{b}$ that describe the atomic system. The operatorial nature of these equations makes the statistical properties of the modes $\hat{a}_n$ rich and relatively hard to extract. In particular, obtaining the Wigner function as in Eqs. (S54-S57) is numerically rather challenging, and subject to rounding errors that become severe when the involved number of photons is large. In this Section, we propose an alternative approach based on a phase space technique that, while introducing an approximation, can be very efficient even when the involved number of photons is large (in fact, especially in this case).

## S8.1: General idea

The idea of phase-space methods is to replace the Heisenberg dynamics for the operators $\hat{\boldsymbol{b}}(0)$ by an ensemble of classical trajectories for fields $\boldsymbol{\psi}(0)$. Information on quantum properties of the system (such as correlations and fluctuations) can then be extracted in terms of averages over the classical ensemble [16]. The ensemble of trajectories is obtained integrating classical equations of motion starting from an ensemble of initial conditions, that should be drawn stochastically from a distribution chosen according to the actual (single) quantum initial condition, so to mimic its intrinsic quantum fluctuations. Specifically, in our case we want to generate an ensemble of $R \gg 1$ atomic classical initial conditions $\{\boldsymbol{S}^{(r)}(0)\}$, with $r = 1,2,\dots,R$, and run the atom-photon dynamics for each of them, obtaining ensembles of classical atomic trajectories $\{\boldsymbol{S}^{(r)}(t)\}$ as well as classical photonic fields $\{\alpha_n^{(r)}\}$ for each harmonic $n$ of interest. From the distribution of the complex numbers $\{\alpha_n^{(r)}\}$, we can extract information on the Wigner function, photon statistics, and more. In this Section, we briefly review the fundamentals of phase-space methods, discuss how to sample the ensemble of classical initial conditions for the quantum initial conditions of interest, and show how this methods compares with the results in the main text.

Although phase-space methods have been devised for a multitude of systems [16,17], their biggest success has arguably been achieved for bosonic systems, for which the most popular method takes the name of "Truncated Wigner Approximation (TWA)". Indeed, the core idea of phase-space methods is to add quantum fluctuations on top of a leading classical dynamics, and bosonic systems have the advantage of featuring a clear notion of classical (mean-field) limit in which TWA becomes more and more accurate. This limit is that of large occupations numbers of the bosonic modes, in which complex fields $(\beta_j, \beta_j^*)$ can effectively play the role of the bosonic modes $(\hat{b}_j, \hat{b}_j^\dagger)$. The TWA for bosonic systems is a well-established technique, and its derivation, limits of validity, and applications are known [18]. For instance, it was shown in [19] that TWA leads to an error $\sim 1/\rho$, where $\rho$ is the density of bosons in the modes. As we have shown in Section S2, the system of interest here is *effectively fully bosonic*: the photons are intrinsically bosonic, whereas the atoms can be mapped to a system of bosonic modes $\hat{\boldsymbol{b}}$. On the one hand, if the emission is strong enough the photonic modes can eventually become highly populated. On the other hand, the first two modes $(\hat{b}_1, \hat{b}_1^\dagger)$ and $(\hat{b}_2, \hat{b}_2^\dagger)$, that we assumed to be the only ones populated at initial time and therefore the most relevant, are highly populated if $N \gg 1$. Therefore, indeed, our system of interest is well suited for treatment within a phase space method.

## S8.2: Truncated Wigner Approximation – key aspects

The TWA is a method devised for dealing with the dynamics of bosons on a lattice. We shell first very quickly review it in its "traditional" form (although we refer the reader to [16,18] for thorough and more accurate reviews), and then revise it to adapt it to our system and needs.

*Fundamentals of "traditional" TWA:*

Quoting from [20], when dealing with bosons in a lattice, "the whole idea of the TWA is that the expectation value of any given operator $\Omega$ at time t is equal to the corresponding classical observable $\Omega_{cl}(t)$ evaluated according to standard Gross-Pitaevskii Equation (GPE) and averaged over an ensemble of initial conditions distributed according to the Wigner transform of the initial density matrix

$$\langle F(t) \rangle = \int d\psi_0^* d\psi_0 W(\psi_0, \psi_0^*) F_{cl}(\psi(t), \psi^*(t))" \tag{S70}$$

(change of notation ours). Let us make a few remarks on Eq. (S70). Therein, $W(\psi_0, \psi_0^*)$ is the Wigner function corresponding to the (single) quantum initial condition, whereas $\psi$ denotes a vector of complex fields, $\psi = (\psi_1, \psi_2, \dots, \psi_L)$. Specifically, the complex field $\sqrt{N}\psi_j$ is the classical variable playing the part of the bosonic mode $\hat{c}_j$ (whereas $\hat{c}_j^\dagger \to \sqrt{N}\psi_j^*$), assuming $L$ modes in the system and $N$ particles. For a given classical initial condition $\psi_0$ of the fields $\psi$, the classical dynamics $\psi(t)$ is obtained from the GPE, that is nothing but the classical version of the Heisenberg dynamical equation for the bosonic modes $\hat{c} = (\hat{c}_1, \hat{c}_2, \dots, \hat{c}_L)$ obtained upon replacing $\hat{c} \to \sqrt{N}\psi$. Given an observable $F = f(\hat{c}, \hat{c}^\dagger)$ that is a function of the bosonic modes, its classical counterpart is obtained replacing the modes with the complex fields, that is, $F_{cl} = f(\psi, \psi^*)$. Note, the ambiguity on the operator ordering in $f$ does not matter, since it falls anyway within the error $\sim 1/N$ of the method.

The integral in Eq. (S70) can be very hard to compute, especially if the number of involved modes $L$ is large. In practice, integrals in high dimension (note, the integration variable in Eq. (S70) has dimension $2L$) can be computed with Monte Carlo methods. In essence, these methods individuate a positive definite part of the integrand, use it as a probability distribution to draw points in the high dimensional space, and compute the integral as an ensemble average over the points of the remaining part of the integrand. Unfortunately, however, no such a positive definite part appears explicitly in the integrand of Eq. (S70). In fact, what would have appeared as the most natural choice, the Wigner function $W$, is a *quasi*-probability and can thus take negative values. To overcome this obstacle, when adopting the TWA it is standard to replace the non-positive-definite (and normalized) quasi-probability distribution $W$ with a positive-definite (and still normalized) probability distribution $p$. At the price of introducing an approximation, this allows the replacement à la Monte Carlo of the integral $\int d\psi_0^* d\psi_0 W(\psi_0, \psi_0^*)$ with an ensemble average,

$$\langle F(t) \rangle = \frac{1}{R} \sum_{r=1}^{R} F_{cl}\left(\psi^{(r)}(t), \psi^{(r)*}(t)\right), \tag{S71}$$

with $\{\psi^{(r)}(t)\}$ the fields obtained from the classical (GPE) evolution of $\{\psi_0^{(r)}\}$, the ensemble of classical initial fields drawn at random according to the distribution $p(\psi_0, \psi_0^*)$, and $R \gg 1$ the number of classical trajectories (that should be taken to be large enough so to observe convergence of the results).

*TWA revised:*

We are now interested in understanding how a phase space method akin to TWA can be used in our setting. If on the one hand our model may appear unconventional from a condensed matter perspective (e.g., as compared to standard Bose Hubbard models [21], on the other hand it is also true that the idea of phase-space methods to describe quantum systems with an ensemble of classical trajectories is all but new in the context of quantum optics [22], and is also specifically used in the context of superradiance [9]. Using the language of TWA flashed in the previous paragraphs, our system can be fully described in terms of bosonic modes $\hat{c} = (\hat{b}_1, \hat{b}_2, \hat{a}_1, \hat{a}_3, \hat{a}_5, \dots, \hat{a}_n, \dots)$, of which the first two describe the collective atomic spin, and the remaining describe the modes of the harmonics. The respective classical complex fields are $\psi = (\beta_1, \beta_2, \alpha_1, \alpha_3, \alpha_5, \ldots, \alpha_n, \ldots)$. The idea for a phase space method is then the following:

1. Generate an ensemble of classical atomic initial conditions $\{\mathbf{S}^{(r)}(0)\}$, with $r = 1, 2, \ldots, R$. This ensemble corresponds to an ensemble of bosonic fields $\{\beta_1^{(r)}(0), \beta_2^{(r)}(0)\}$ through the mapping $\mathbf{S} = (\beta_1, \beta_2)\boldsymbol{\sigma}\begin{pmatrix}\beta_1^*\\\beta_2^*\end{pmatrix}$. Details on how this sampling is done will be provided in the next subsection;

2. Use Eq. (S30) to solve the classical trajectories, by just replacing $\widehat{\boldsymbol{b}}(t) \to \boldsymbol{\beta}^{(r)}(t)$ therein. In this way, we have access to the classical dynamics of all the $R$ trajectories: $\boldsymbol{\beta}^{(r)}(t) = \boldsymbol{F}(t)\boldsymbol{\beta}^{(r)}(t)$, where the transition matrix $\boldsymbol{F}(t)$ has to be computed once and for all (and contains information on the single particle atomic spectrum, see Sections S1 and S4);

3. Plug the classical trajectories $\{\boldsymbol{\beta}^{(r)}(t)\}$ into the classical version of Eq. (S27) (in which operators are substituted by classical fields) to obtain the respective ensembles of photonic classical fields $\{\alpha_{k\sigma}^{(r)}\}$. The same machinery of the quantum theory in Section S4 applies, leading to an ensemble of classical harmonic fields as in Eq. (7) in the main:

$$\alpha_n^{(r)} = \alpha_n^{(r)}(0) + \alpha + (\boldsymbol{u}_n + i\boldsymbol{v}_n) \cdot \mathbf{S}^{(r)}(0), \tag{S71}$$

In the latter, $\mathbf{S}^{(r)}(0)$ are the atomic initial conditions in point (1) above, whereas $\alpha_n^{(r)}(0)$ has normally distributed real and imaginary parts with standard deviation $1/\sqrt{2}$, to reflect vacuum fluctuations (indeed the Wigner function of the vacuum is $W_{vac}(X, P) = \frac{1}{\pi}e^{-X^2-P^2}$);

4. Compute an ensemble of classical photon numbers as $k_n^{(r)} = round\left(\left|\alpha_n^{(r)}\right|^2\right)$;

5. Obtain the photon statistics as the statistics of the classical photon numbers $\{k_n^{(r)}\}$.

6. Compute the classical Wigner function as a two-dimensional histogram in the complex plain of the classical fields $\{\alpha_n^{(r)}\}$. In practice, rather than computing such a histogram, we prefer to plot $\{\alpha_n^{(r)}\}$ with a scatter plot in the complex plain, imprinting the density of points in their colorcode. As an important limitation, however, the method systematically misses any possible negativity of the Wigner function.

We note that points 5 and 6 are particularly original, in the sense that TWA is normally devised to compute expectation values of some observables, rather than the photon statistics or Wigner function associated to a mode. The validity of this approach will be validated in the following by a direct comparison of the above described phase space approach and fully quantum theory.

### S8.3: Sampling the classical initial conditions

We now turn to how the ensemble of initial conditions should be sampled. The goal is to find a distribution that resembles the atomic Wigner function (at time 0) as much as possible, but is positively definite.

To begin with, we compute the atomic Wigner function on the Bloch sphere for selected atomic states of interest, that is, for the state with all excited atoms $|\Uparrow\rangle$ and for that with only half of the atoms excited, $|N/2\rangle$. Both states are invariant under any permutation of the atoms, and their Wigner function $W(\theta, \phi)$ on the Bloch sphere, with $\theta$ and $\phi$ the polar and azimuthal angles, respectively, can be computed as in [14]. The symmetry of $|\Uparrow\rangle$ and $|N/2\rangle$ makes their Wigner $W(\theta, \phi)$ independent of the azimuthal angle $\phi$, making it possible for these states to only reason in terms of quasiprobability of the polar angle $\theta$. Considering the Jacobian factor, we call this $w(\theta) \propto |\sin\theta| W(\theta, \phi)$, and plot it in Figure S6. The goal is to replace the quasi-probability $w(\theta)$ with a positive definite probability distribution $p(\theta)$. We find that a good functional form for $p(\theta)$ is the following

$$p(\theta) \propto |\sin\theta| e^{-\frac{(\theta-\theta_0)^2}{2\sigma^2}}, \tag{S72}$$

where $\theta_0 = 0, \pi/2$ for $|\Uparrow\rangle$ and $|N/2\rangle$, respectively, and in which the width $\sigma$ is found by directly fitting the quasiprobability $w(\theta)$. In the case of $|\Uparrow\rangle$ (left in Figure S7), the replacement $w(\theta) \to p(\theta)$ appears very natural and accurate, because both $w(\theta)$ and $p(\theta)$ are positive definite. In the case of $|N/2\rangle$ (center in Figure S7), $w(\theta)$ oscillates taking both positive and negative values. In this case, the replacement with $p(\theta)$ in Eq. (S72) is motivated by the fact that the fast oscillations away from the main central peak at $\theta = \pi/2$ compensate one another in Eq. (S70) [20], and can thus be effectively neglected. In the right panel in Figure S7, by repeating the above procedure for various number of atoms $N$, we find that the width $\sigma$ in Eq. (S72) scales with $N$ as $\sigma = a_1 N^{-a_2}$, the coefficients $a_1$ and $a_2$ being given by a direct fit, and found to be $a_1 = 0.8887$ and $a_2 = 0.4741$ for $|\Uparrow\rangle$, and $a_1 = 0.8956$ and $a_2 = 0.9549$ for $|N/2\rangle$. By simple extrapolation, we can in this way get $\sigma$, and thus the positive definite distribution $p(\theta)$ in Eq. (S72) for any $N$.

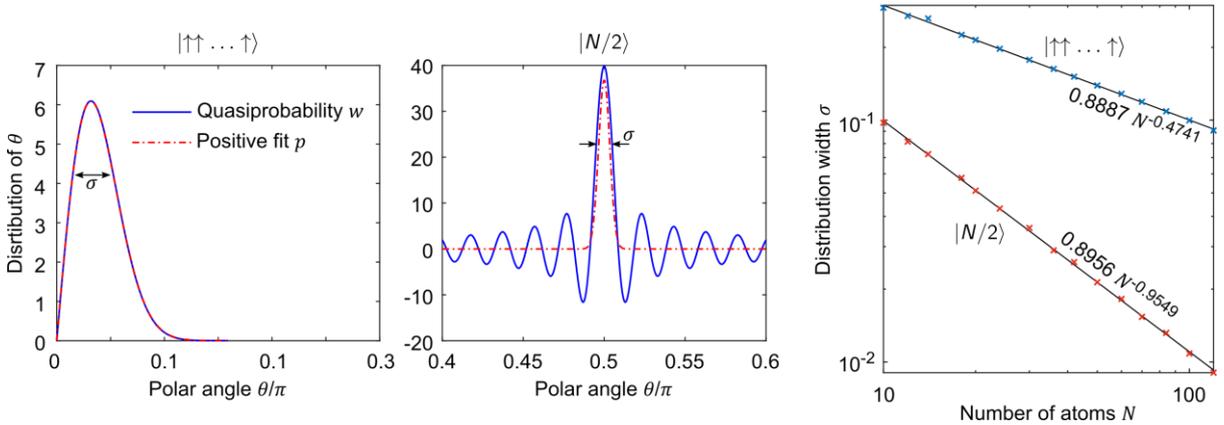

**Figure S7. Analysis of the atomic Wigner function and its numerical fitting**, for the cases of all-excited atoms $|\uparrow\uparrow\cdots\uparrow\rangle$ and half-excited atoms $|N/2\rangle$. For $|\uparrow\uparrow\cdots\uparrow\rangle$ and $N = 100$ (left), the Wigner function $w(\theta)$, including the Jacobian factor $\sin\theta$, is positive. For $|N/2\rangle$ and $N = 100$ (center), the Wigner function features instead a main peak at $\theta = \pi/2$ (the equator of the Bloch sphere) and rapid oscillations around it. Positive fits according to the expression in Eq. (S73) are shown in red. The fit works excellently for $|\uparrow\uparrow\cdots\uparrow\rangle$, and gets rid of the fast oscillations for $|N/2\rangle$. We find the fitting parameter $\sigma$ for various system sizes $N$, and plot it in logarithmic scale (right) for $|\uparrow\uparrow\cdots\uparrow\rangle$ (red) and $|N/2\rangle$ (blue). This highlights a scaling $\sigma = a_1 N^{-a_2}$, with coefficients $a_{1,2}$ depending on the atomic system state ($a_{1,2}$ are found through direct interpolation).

The distribution $p(\theta)$, now at hand for any system size $N$ and for the states $|\Uparrow\rangle$ and $|N/2\rangle$, can thus be used to sample the ensemble of classical initial conditions on the Bloch sphere according to

$$\theta^{(r)} \sim p(\theta), \qquad \phi^{(r)} = \text{randu}(0, 2\pi). \tag{S73}$$

That is, $\theta^{(r)}$ is distributed according to $p(\theta)$, for which a standard rejection method can be used, whereas $\phi^{(r)}$ is distributed uniformly between 0 and $2\pi$. Here we have discussed only the cases of $|\Uparrow\rangle$ and $|N/2\rangle$, but the discussion can be readily extended to other atomic initial conditions. For instance, the ensemble corresponding to $|\Downarrow\rangle$ can be obtained from that associated to $|\Uparrow\rangle$ by simply adding $\pi$ to all the polar angles. From the ensemble of polar coordinates, we can then obtain the ensemble of points on the Bloch sphere $\{\boldsymbol{S}^{(r)}(0)\}$, and therefore that of bosonic fields $\{\boldsymbol{\beta}^{(r)}(0)\}$.

### S8.4: Results

The above prescription allows us to quickly obtain information on the properties of the quantum emission. We show this by reproducing, in Figure S8, Figure 2 from the main. Indeed, the two are qualitatively very similar. While the phase space results in Figure S8 are more noisy and less precise than those in Figure 2, they are also much easier to obtain. Indeed, the computation of the Wigner function as in Figure 2 can be numerically very challenging, whereas the results in Figure S8 are obtained by simply generating an ensemble of initial conditions and by mapping them to the complex plane of the photons (see point (3) in Section S8.2 above). Moreover, the phase space approach and Eq. (S71) in particular offer a clear insight into the shape of the Wigner function of the output harmonics. Indeed, Eq. (S71) provides a direct link between the atomic initial conditions on the Bloch sphere $\boldsymbol{S}^{(r)}(0)$ and the output photonic fields $\alpha_n^{(r)}$. The link unfolds as follows: in Eq. (S71), the term $(\boldsymbol{u}_n + i\boldsymbol{v}_n) \cdot \boldsymbol{S}^{(r)}(0)$ corresponds to a projection of the points on the sphere onto the plane individuated by the vectors $\boldsymbol{u}_n$ and $\boldsymbol{v}_n$. Such plane might also get stretched due to the angle between $\boldsymbol{u}_n$ and $\boldsymbol{v}_n$, and their different lengths. The term $\alpha$ represents then a shift in the complex plane, whereas $\alpha_n^{(r)}(0)$ adds a Gaussian noise to each point. With this in mind, we can therefore intuitively predict, for a given $\boldsymbol{u}_n$, $\boldsymbol{v}_n$, and initial atomic condition, the qualitative shape of the output Wigner function.

We conclude with a few remarks. First, we note that this phase space method exploits lots of the machinery used for the fully quantum theory described above and adopted in the main text. Indeed, the phase space approximation essentially only enters at the level of the final equation of the theory (Eq. (7) in the main), where the initial atomic condition is substituted by an ensemble of classical initial conditions. Therefore, in particular, the theory still exploits and requires the computation of the $2 \times 2$ dimensional matrix $\boldsymbol{d}_n$. Second, we note that the phase space method can be readily adapted to account for a preparation stage: indeed, both the protocols that we considered, that based on superradiance and that based on the interatomic interactions, can be simulated within the phase space method by just integrating the respective classical atomic dynamical equations for the each classical trajectory. Third and last, we observe that the phase space method can also provide information on the correlation between harmonics. Indeed, by computing the classical complex light fields $\{\alpha_n^{(r)}, \alpha_m^{(r)}\}$ for two harmonics $n$ and $m$, and drawing them as points in the plane of, say, the real part of $\alpha_n^{(r)}$ and the imaginary part of $\alpha_m^{(r)}$, one can readily access information on the joint Wigner function of the two harmonics, which in the fully quantum treatment would be much more challenging.

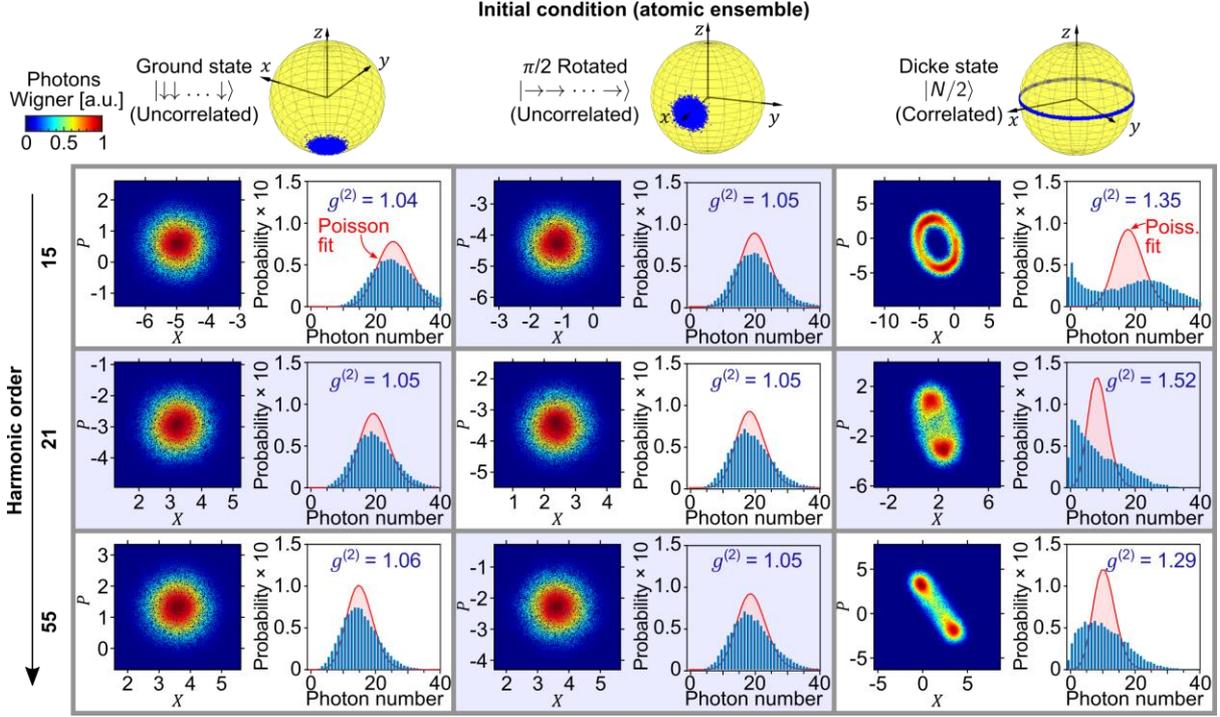

**Figure S8. Many-body high-harmonic generation within a phase space method.** We reproduce Figure 2 from the main within a phase space approach. The parameters considered here are the same as in Figure 2 in the main, and $R = 20000$ trajectories are considered. On top, the atomic Wigner function is replaced by a scatter plot of the atomic initial conditions on the Bloch sphere. The photonic Wigner function is instead obtained as a scatter plot of the ensemble $\{\alpha_n^{(r)}\}$ in the complex plane, the color of the dots representing their density. A good qualitative agreement of the plots with those in Figure 2 is apparent.